\documentstyle[epsf]{espart}
\newcommand{\gb}[1]{\gamma^{\left({#1}\right)}}
\newcommand{\gbb}[1]{
\left(\gamma^{\left({#1}\right)}\otimes\gamma^{\left({#1}\right)}\right)}
\newcommand{\N}[1]{N\left[{#1}\right]}
\newcommand{\Np}[1]{N^\prime\left[{#1}\right]}
\begin{document}
%
%
%
%
\begin{frontmatter}
\hfill IFUP-TH 9/94

\hfill HUTP-94/A001

\title{QCD corrections to electroweak processes in an unconventional
scheme: application to the $b \rightarrow s \gamma$ decay}
\author[IFUP]{Giancarlo~Cella\thanksref{MURST}},
\author[INFN]{Giuseppe~Curci},
\author[HARVARD]{Giulia~Ricciardi\thanksref{INFNr}} and
\author[INFN]{Andrea~Vicer\'e\thanksref{INFNv}}
\address[IFUP]{Dipartimento di Fisica dell'Universit\`{a} di Pisa, Piazza
Torricelli 2, I-56126 Pisa, ITALY,  e-mail {\tt cella@sun10.difi.unipi.it}}
\address[INFN]{Istituto Nazionale di Fisica Nucleare, Piazza
Torricelli 2, I-56126 Pisa, ITALY, e-mail {\tt vicere@sun10.difi.unipi.it}}
\address[HARVARD]{Lyman Laboratory of Physics, Harvard University,
Cambridge, MA 02138, USA, e-mail {\tt ricciardi@physics.harvard.edu}} 

\thanks[MURST]{ supported in part by M.U.R.S.T., Italy}
\thanks[INFNr]{supported in part  by I.N.F.N., Italy
and by the National Science Foundation under Grant No. PHY-92-18167}
\thanks[INFNv]{supported in part  by I.N.F.N., Italy}

\begin{abstract}
In this work we give a detailed description of a method for the
calculation of QCD corrections to electroweak processes
in dimensional regularization that does not
require any definition of the $\gamma_5$ matrix
in $d$ dimensions. 
This method appears particularly convenient in order to limit the
algebraic complexity of higher order calculations.
As an example, we compute the leading logarithmic corrections
to the $b\rightarrow s \gamma$ decay.
\end{abstract}
\end{frontmatter}
\newpage
%
%
\section{Introduction}
\label{sec:introduction}
In the Standard Model the flavor changing neutral current (FCNC)
processes can take place only as loop effects because of the GIM
mechanism. 
Therefore they are strongly suppressed compared with ordinary weak
decays and very sensitive to the occurrence of new physics. Among all
FCNC processes, the inclusive decay $B \rightarrow X_s \gamma$ is
particularly interesting. In fact, due to the relatively heavy mass
of the b-quark ($m_b \gg \Lambda_{QCD}$), the long distance QCD
contributions are expected to play a minor role and the process can be
confidently modeled on the $b \rightarrow s \gamma$ decay in the 
spectator model, corrected for the short distance QCD corrections. 

This approach has allowed a clean theoretical prediction of the inclusive
rate, which is compatible with the recent experimental results from the CLEO
collaboration~\cite{cleo:prl:71/93,ams:thorndike:4/93}. For instance
at $m_t = 174 \pm 16\,{\mathrm GeV}$~\cite{CDF:94} the theoretical prediction
\begin{equation}
BR\left[B \rightarrow X_s\gamma\right]_{\mathrm theor} \pm \Delta_{\mathrm
theor} \pm \Delta_{m_{top}}= \left(3.0 \pm 0.8 \pm 0.1 \right)\, 10^{-4}
\label{eq:tp}
\end{equation}
is larger than the exclusive fraction $B\rightarrow K^\star\gamma$ 
\begin{equation}
BR\left[B \rightarrow K^\star \, \gamma\right]
=(4.5\pm 1.5\pm0.9)\, 10^{-5}
\end{equation}
and smaller than the measured lower bound (at 95\% confidence level) on the
inclusive rate
\begin{equation}
BR[B \rightarrow X_s \, \gamma]
< 5.4\, 10^{-4}.
\end{equation}
Presently only the leading order QCD corrections (LO) are known.
The importance of the next to leading order (NLO) corrections has been
recently stressed by Buras et al.~\cite{buras:TUM-T31-50/93}. They have
shown that most of the uncertainty in Eq.~(\ref{eq:tp}) comes from the
residual dependence on the scale of the decay, and the inclusion of the
subleading logarithms is a necessary step toward the reduction of the
theoretical error, from about 25\% to less  than 10\%.

It goes without saying that any search for new physics through the window
of the rare $B$ decays  is seriously hampered by so large an uncertainty in
the Standard Model predictions. Since we expect that perturbative QCD works
well at this scales of energy, at least for the inclusive predictions, the
effort needed to reduce this uncertainty seems worthwhile.

In this paper we give full details about the computation of the LO
corrections to $b \rightarrow s \gamma $ decay
in dimensional regularization; the results of this work
have already been published in~\cite{CCRV:letter}.
We have applied to this problem a scheme that we consider an important step
toward the computation of the NLO corrections and we want to substantiate
the statement that this is a practical alternative to the t'Hooft-Veltman
scheme, for this particular class of computations and in the context of
dimensionally regularized field theory.
To be definite, the method is helpful when evaluating the short
distance corrections to a weak effective hamiltonian, induced by a theory
that is invariant under the transformation $\gamma_5\rightarrow -
\gamma_5$, like perturbative QCD. 
This scheme has been introduced in~\cite{curci:ricc:prd47} and
successfully applied to the NLO QCD corrections of a generic $\Delta
F=1$ process in the absence of penguin operators.

Here we describe this method by
applying it to the LO
$b \rightarrow s \gamma$
computation. 
Our conclusions are that the latest LO existing results, obtained by
Ciuchini et al.~\cite{ciuchini:plb316,ciuchini:rome93/973}, are correctly
reproduced by this method and that it seems there are no
technical obstructions to the computation of QCD corrections at all
orders in perturbation theory.

We find useful to sketch here some major steps in the history of the
LO corrections. The big enhancement of the decay rate due to QCD
corrections was pointed out
in~\cite{shifman:PRD18,bertolini:PRD59,deshpande:PRL59}, in the case of a
light top quark.
The first works where the top mass was considered sizable showed a
substantial disagreement.
In particular the more complete works of Grinstein et
al.~\cite{grinstein:plb202,grinstein:npb339,grinstein:npb365} and 
Grigjanis et al.~\cite{grigjanis:plb213,grigjanis:plb224} seemed to
show a contradiction between the naive dimensional regularization
(NDR), based on the use of an anticommuting $\gamma_5$ matrix in $d$
dimensions and the method of dimensional reduction
(DRED)~\cite{siegel}.

This mismatch was practically very relevant: for example at
$m_{t} = 140 \,\,\mathrm{GeV}$ the enhancement of the QCD corrected
branching ratio was about a factor $4$ in the NDR scheme, to be
compared with a factor $2$ in DRED. 

In our work~\cite{CCRV:plb248} we confirmed the result of the NDR
analysis.  All these computations were done in a reduced basis, not
including certain operators whose contribution was
supposed to be small~\cite{grinstein:npb339}.

The computation was extended to the complete basis by
Misiak~\cite{misiak:plb269,misiak:npb393}; the whole problem was
reconsidered recently by Ciuchini et
al.~\cite{ciuchini:plb316,ciuchini:rome93/973}, who have solved the 
contradiction between the NDR and DRED schemes demonstrating that the
anomalous dimension (AD) matrix, even at leading logarithmic order, is
scheme dependent. This scheme dependence is canceled in the
physical amplitude just by the matrix elements of the operators left
out in the first computations. Misiak has further observed
in~\cite{misiak:TUM-T31-46/93} that the scheme dependence could have
been partly compensated  in the reduced basis used by us
in~\cite{CCRV:plb248} by considering some matrix elements of
operators related by the motion equations to the terms left out in the
complete basis. 

The results for the AD matrix reported
in~\cite{ciuchini:plb316,ciuchini:rome93/973} and
in~\cite{misiak:npb393} are in slight disagreement, although this
difference has little relevance on the phenomenology. In the present paper
we confirm the AD matrix presented
in~\cite{ciuchini:plb316,ciuchini:rome93/973}. 

The plan of the paper is as follows: in Section~\ref{sec:strategy} we
describe the method on general grounds and we show that the
computation of the anomalous dimension matrix can be
considerably simplified by splitting the effective hamiltonian in
parts even and odd under the $\gamma_5 \rightarrow - \gamma_5$
transformation. After an extension of the $4$ dimensional
operators to $d$ dimensions we can avoid the use of a $4$
dimensional $\gamma_5$, which is required in the t'Hooft-Veltman
scheme~\cite{tHooftVeltman}, without any loss of mathematical
rigor. In Section~\ref{sec:bsgamma} we describe the computation of the QCD
corrections to the $B\rightarrow X_s\gamma$ decay: while in our
previous letter~\cite{CCRV:letter} we have given the results in an
on-shell basis, here we consider useful to detail the work in an
off-shell basis; the reason is that the intermediate results,
although redundant, are useful for the planned next to leading
computation. In section~\ref{sec:checks} we comment on the
simplifications characteristics of this scheme, particularly the 
possibility to use generalized Fierz identities~\cite{avdeev} to
relate Feynman diagrams with and without fermionic traces: the
existence of charge conjugation properties further reduces the number
of independent graphs.

The appendixes are devoted to reference formulas: in App.~\ref{app:rg}
we list some results on the renormalization group in  the context of
dimensional regularization and we comment on the  treatment of
evanescent operators. In App.~\ref{app:dirac} we give a set  of
formulas needed to implement the Dirac algebra: they are known in  
literature, but we think useful to collect them in a ``ready-to-use''
form. 
%
%
\section{Strategy}
\label{sec:strategy}
In the following, we will explain our approach to the computation of
the RG evolution. After reminding briefly the general framework for
the computation of QCD corrections to weak decays, we will give full
details about the symmetrized scheme we use~\cite{curci:ricc:prd47}.

The starting point is a ``complete'' theory of weak interactions and
 QCD, for instance the Standard Model, which enables us to predict
amplitudes between initial and final states, at some order in
perturbation theory
\begin{equation} 
\label{eq:fullamplitude}
\left<f|i\right>_{\mathrm{complete}} .
\end{equation}
It is well known that it is possible to simplify the computation
of the amplitude in Eq.~(\ref{eq:fullamplitude}) building an effective
hamiltonian ${\cal H}_{\mathrm{eff}}$ that models the effect of the
degrees of freedom which are ``heavy'' compared to the typical energy
scale of the process, in the sense that
\begin{equation}
  \left<f|i\right>_{\mathrm{complete}} = \left<f\left|{\cal
    H}_{\mathrm{eff}}\right|i\right>_{\mathrm{light}} +
  O\left(\frac{1}{M^4}\right)\ ,
\label{eq:expansion}
\end{equation}
where $M$ is the ``heavy'' scale.
${\cal H}_{\mathrm{eff}}$ is a sum of local operators built of
``light'' fields. The dynamics is specified by a reduced lagrangian
with the heavy fields deleted and with couplings and masses 
which undergo a finite renormalization while passing from the
``complete'' to the effective theory. 

The effective hamiltonian is determined via a matching procedure: the
operator content of ${\cal H}_{\mathrm{eff}}$ is obtained with considerations
on the residual symmetries of the reduced lagrangian and the coefficients are
computed imposing Eq.~(\ref{eq:expansion}) on a finite number of processes.
An important point is that, even when the ``complete'' theory is
finite (for instance, when the divergences are canceled by the 
Glashow-Iliopoulos-Maiani mechanism), the effective hamiltonian
contains operators, weighted with some inverse power of $M$, whose
matrix elements are individually divergent and must be renormalized.
A renormalization scale $\mu$ must therefore be introduced; we can
write the effective hamiltonian as
\begin{equation}
{\cal H}_{\mathrm{eff}} = \frac{1}{M^2}\sum_i C_i\left(\mu,\,M\right)
\N{O_i}\ ,
\end{equation}
where the $N$ operation is a subtraction procedure, for instance
minimal subtraction in dimensional regularization~\cite{bonneau}.
The matrix elements between on-shell states,
\begin{equation}
\left<f\left|\N{O_i}\right|i\right>\left(\mu\simeq M\right)
\end{equation}
contain large perturbative contributions depending on $\log{\mu/m}$,
where $m$ is the typical scale of the external states.  These large
logarithms can be summed by exploiting the renormalization group
invariance of the combination
\begin{equation}
  \sum_i C_i\left(\mu,\,M\right)
  \left<f\left|\N{O_i}\right|i\right>\left(\mu\right).
\end{equation}
By scaling down the whole expression to values of $\mu\simeq m$, the
large logarithms are transferred in the coefficients and the matrix
elements can be computed more reliably.

The standard way to determine the RG evolution of the matrix elements
is the computation of their anomalous dimension matrix $\hat{\gamma}$,
which appears in the RG equation
\begin{equation}
  \mu \frac{\d}{\d \mu} \N{O_i} + \gamma_{ij} \N{O_j} = 0\ .
\end{equation}
While the transition amplitude has physical meaning, the three steps
of the computation, the matching at the $\mu\simeq M$ 
scale, the determination of the AD matrix and the evaluation of the
matrix elements at the $\mu\simeq m$ scale, are in general all scheme
dependent: we want to choose a framework which simplifies the most
difficult part, the computation of the $\hat{\gamma}$ matrix.

\subsection{An unconventional approach}
\label{subsec:thescheme}

Let us assume to have defined the effective theory in a regularization
scheme free from ambiguities, like the t'Hooft-Veltman scheme, and to
deal with a certain set of relevant operators, that is, operators with
a non-zero naive limit in $4$-dimensions. They can
also appear evanescent operators (classically zero in
$4$-dimensions) that can be ``reduced'' on the relevant ones, and
decoupled at the level of the RG evolution, as we shall see.

Let us consider the general structure of the effective 
hamiltonian, without specifying
the field content
\begin{eqnarray}
\label{eq:heff}
{\cal H}_{\mathrm{eff}} &=& \sum_i C^{\mathrm{R,HV}}_i \N{R_i} +
\sum_i C^{\mathrm{L,HV}}_i \N{L_i}
\nonumber\\ &+& \sum_i
C^{\mathrm{LL,HV}}_i \N{\left(L\otimes L\right)_i} + \sum_i
C^{\mathrm{LR,HV}}_i \N{\left(L\otimes R\right)_i}\ .
\end{eqnarray}
The $\N{L}$ and $\N{R}$ operators are bilinear in the Fermi fields, like
for instance $O_2$ and $O_5$ in the basis in Eq.~(\ref{primabase}), while
$\N{L\otimes R}$ and $\N{L\otimes L}$ stand for current-current
operators, like $O_{11}$ and $O_{13}$.
The symbols $R$ and $L$ remind the presence of the chiral projectors
\mbox{$P_{\frac{L}{R}} = \frac{1}{2} (1 \mp \gamma_5)$}, where
\mbox{$\gamma_5 = -\gamma_1\gamma_2\gamma_3\gamma_4$} in euclidean notation.
Now if we ``redefine'' the theory with the substitutions
\begin{eqnarray}
\label{eq:redefine}
\gamma_5 & \rightarrow & \bar{\gamma}_5 = - \gamma_5 \nonumber \\
\epsilon_{\mu \nu \rho \sigma} & \rightarrow & \bar{\epsilon}_{\mu \nu
  \rho \sigma} = - \epsilon_{\mu \nu \rho \sigma}
\end{eqnarray}
and compute the effective hamiltonian with the new Feynman rules
(and the same regularization and renormalization scheme) we obtain
an expression identical to the one in Eq.~(\ref{eq:heff}), except for the
substitutions in Eq.~(\ref{eq:redefine}). In fact, the commutation and
trace rules are invariant under the transformation in
Eq.~(\ref{eq:redefine}), for example ($\tilde{\gamma}_\mu$ is the
Dirac matrix in the $d-4$ unphysical space)
\begin{eqnarray}
\left[\gamma_\mu,\gamma_5\right]_+ = 2 \gamma_5 \tilde{\gamma}_\mu &
\rightarrow & \left[\gamma_\mu,\bar{\gamma}_5\right]_+ = 2 \bar{\gamma}_5
\tilde{\gamma}_\mu\nonumber\\
\mathrm{Tr}\left(\gamma_5 \gamma_\mu \gamma_\nu \gamma_\rho
\gamma_\sigma\right) =
4 \epsilon_{\mu \nu \rho \sigma} & \rightarrow & 
\mathrm{Tr}\left(\bar{\gamma}_5 \gamma_\mu \gamma_\nu \gamma_\rho
\gamma_\sigma\right) = 4 \bar{\epsilon}_{\mu \nu \rho \sigma}\ .
\end{eqnarray}
With the same argument it is easy to conclude that the RG
evolution equation is conserved by the transformation~(\ref{eq:redefine})
\begin{eqnarray}
\label{eq:rgtwo}
\left[ \mu \frac{\d}{\d \mu} + \hat{\gamma} \right]
\left(
\begin{array}{c}
  \N{L \otimes L} \\ \N{L \otimes R} \\ \N{R} \\ \N{L}
\end{array} 
\right)
= 0 
& ~\mapsto~ &
\left[ \mu \frac{\d}{\d \mu} + \hat{\gamma} \right]
\left(
\begin{array}{c}
  \N{R \otimes R} \\ \N{R \otimes L} \\ \N{L} \\ \N{R}
\end{array} 
\right)
= 0\ .
\end{eqnarray}
The same anomalous dimension matrix $\hat{\gamma}$ appears in both the
equations in~(\ref{eq:rgtwo}).
The key point is that we are interested to the RG evolution
determined only by QCD on a given set of operators.
But the QCD is left unchanged by the
transformation in Eq.~(\ref{eq:redefine}), because gluons have a
purely vectorial coupling, so the two relations in~(\ref{eq:rgtwo})
are simultaneously true.
It follows that the RG evolution is the same for every linear
combination of the two operator sets, and for the purpose of the
calculation of the AD matrix all these combinations are completely
equivalent. It will be convenient in particular to consider the RG
evolution of the symmetric combination,
\begin{eqnarray}
\left[ \mu \frac{\d}{\d \mu} + \hat{\gamma} \right]
\left(
\begin{array}{c}
  \frac{1}{2} \left( \N{L \otimes L} + \N{R \otimes R} \right) \\ 
  \frac{1}{2} \left( \N{L \otimes R} + \N{R \otimes L} \right) \\ 
  \frac{1}{2} \left( \N{R} + \N{L} \right)
\end{array} 
\right)
= 0,
\end{eqnarray}
because it is possible to redefine these operators in order to make
the $\gamma_5$ matrix disappear. There is an automatic cancellation
for the operator $N[R]+N[L]$, as one can easily
see, while for the $4$-fermion operators a change of the $d$
dimensional extension is needed.
A complete (infinite) basis for the Clifford algebra in $d$
dimensions is given by the completely antisymmetric products of
$\gamma$ matrices~\cite{avdeev,kennedy,curci:ricc:prd47},
\begin{eqnarray}
\gb{n} & \equiv & \gamma_{\mu_1,\mu_2,\dots\mu_n} = \frac{1}{n!}
\sum_{p\in \Pi_n} (-1)^p \gamma_{\mu_1}
\gamma_{\mu_2}\dots\gamma_{\mu_n}\, \nonumber \\
\gb{n} \otimes \gb{n} & \equiv & \sum_{\mu_1\dots\mu_n}
\gamma_{\mu_1\dots\mu_n}\otimes \gamma_{\mu_1\dots\mu_n}\ .
\label{eq:defgb}
\end{eqnarray}
In $4$ dimensions the structures $\gb{n}$ survive only for $n\leq 4$ and
one can write
\begin{eqnarray}
\frac{1}{2} \left[ (L \otimes L) + (R \otimes R) \right] & = &
\frac{1}{4} \left[(\gb{1} \otimes \gb{1}) + \frac{1}{3!}(\gb{3}
\otimes \gb{3})\right]
\nonumber \\ 
\frac{1}{2} \left[ (L \otimes R) + (R \otimes L) \right] & = &
\frac{1}{4} \left[(\gb{1} \otimes \gb{1}) - \frac{1}{3!}(\gb{3}
\otimes \gb{3})\right].
\end{eqnarray}
These equations can be taken as {\it definitions} for the tensor
products in $d$ dimensions. 

This is {not equivalent to the t'Hooft Veltman scheme, which is
based on the $\gamma_5$ definition 
\begin{equation}
\gamma_5^{\mathrm{HV}} = - \frac{1}{4!}\,\epsilon_{\mu \nu \rho
\sigma} \gamma_{\mu\nu\rho\sigma}\ .
\end{equation}
By direct substitution, the expansion of the symmetrized scalar
operators in the basis~(\ref{eq:defgb}) is
\begin{eqnarray}
\frac{1}{2} \left( (L \otimes L) + (R \otimes R) \right) & = &
\frac{1}{4} \left(\gamma_\mu \otimes \gamma_\mu  + \gamma_\mu
\gamma_5^{HV} \otimes \gamma_\mu \gamma_5^{HV} \right) \nonumber \\
\frac{1}{2} \left( (L \otimes R) + (R \otimes L) \right) & = &
\frac{1}{4} \left(\gamma_\mu \otimes \gamma_\mu  - \gamma_\mu
\gamma_5^{HV} \otimes \gamma_\mu \gamma_5^{HV} \right)\\
\gamma_\mu \gamma_5^{HV} \otimes \gamma_\mu \gamma_5^{HV} &=&
\frac{1}{3!}\,\delta^4_{\mu_1\mu_2} \delta^4_{\nu_1\nu_2}
\delta^4_{\rho_1\rho_2} \left(\gamma_{\mu_1\nu_1\rho_1}
\otimes\gamma_{\mu_2\nu_2\rho_2}\right)\nonumber\\
&+& \frac{1}{4!} \delta^4_{\nu_1\nu_2}\delta^4_{\rho_1\rho_2}
\delta^4_{\sigma_1\sigma_2}\delta^4_{\tau_1\tau_2}\,
\left(\gamma_{\mu\nu_1\rho_1\sigma_1\tau_1}
\otimes \gamma_{\mu\nu_2\rho_2\sigma_2\tau_2}\right).
\nonumber
\end{eqnarray}
Note that the expansion coefficients are not $O(d)$ invariant
and the symmetrization guarantees only that in each coefficient there
is always an even number of $\epsilon_{\mu\nu\rho\sigma}$ tensors,
that can be reduced to $4$-dimensional $\delta^4_{\mu\nu}$.
On the contrary with our definitions there is no splitting between $4$
dimensional and $d-4$ dimensional objects (the indices in
Eq.~(\ref{eq:defgb}) are $d$ dimensional)
and no breaking of
$O\left(d\right)$ invariance; this will be a
considerable simplification in the following.  For example, as we
work in the well definite basis of $\gbb{n}$ structures, for
arbitrary integer $n$, we are allowed to treat in a unified manner
relevant operators, having $n\leq 4$, and evanescent ones having
$n\geq 5$.
 We shall see that in our case the leading
order anomalous dimension matrix contains two loops terms that are
scheme dependent, so we will not find the same result as in the t'Hooft
Veltman scheme.

We can rewrite the effective hamiltonian in the form
\begin{equation}
{\cal H}_{\mathrm{eff}} = \sum_i C_i^{\mathrm{HV}}\left(\mu\right)
\N{O_i^+}\left(\mu\right) + C_i^{\mathrm{HV}}\left(\mu\right)
\N{O_i^-}\left(\mu\right),
\end{equation}
where $O_i^+$ and $O_i^-$ are even and odd combinations respect
to the transformations in Eq.~(\ref{eq:redefine}).
The two classes of operators are not mixed
by renormalization, hence the two pieces of ${\cal H}_{\mathrm{eff}}$ must
be separately RG invariant
\begin{equation}
  \mu \frac{\d}{\d \mu} \sum_i C_i^{\mathrm{HV}}\left(\mu\right)
\N{O_i^+}\left(\mu\right) = \mu \frac{\d}{\d \mu} \sum_i
C_i^{\mathrm{HV}}\left(\mu\right) \N{O_i^-}\left(\mu\right) = 0\ .
\end{equation}
The redefinition of the symmetric part amounts to add an
evanescent operator $E_i^+$ to each $O_i^+$,
\begin{equation}
  \bar{O}_i^+ = O_i^+ + E_i^+
\end{equation}
and the physics is left unchanged provided the mismatch is reabsorbed in
the coefficients.
This is possible because a renormalized evanescent operator can be
expanded on a complete basis of relevant ones~\cite{collins:book} with
finite expansion coefficients $r_{ij}$
\begin{equation}
  \N{E_i^+} = r_{i j} \N{O_j^+}\ .
\end{equation}
We must impose the condition
\begin{equation}
\sum_i C_i^{HV}\left(\mu\right)
\N{O_i^+}\left(\mu\right) =  \sum_i C_i^{s}\left(\mu\right)
\N{\bar{O}_i^+}\left(\mu\right)
\end{equation}
that gives the result
\begin{equation}
\label{eq:SHVrelation}
C_i^{s}\left(\mu\right) = \left( \delta_{ij} - r_{j i} \right)
C_i^{HV}\left(\mu\right).
\end{equation}
The RG evolution of $C_i^{s}\left(\mu\right)$ coefficients is governed by
the anomalous dimension matrix that is determined by our
procedure and using the Eq.~(\ref{eq:SHVrelation}) it is simple
to find the evolution of $C_i^{HV}\left(\mu\right)$, which is the final aim.
Alternatively the Eq.~(\ref{eq:SHVrelation}) can be interpreted
as a redefinition of the normal product $\N{O}$, with a non-minimal
subtraction procedure;  this is the approach  we will adopt
in this paper.

Now the advantage of this method is apparent: the most difficult part of 
the computation, the determination of the AD matrix, can be done in the 
symmetrized scheme, whose simplicity will be evident in
Sec.~\ref{sec:bsgamma}, while the matching needed to write down the
amplitude requires a computation at one loop order less. 
%
%
\section{The $b\rightarrow s\gamma$ process}
\label{sec:bsgamma}

\subsection{Effective off-shell hamiltonian}
\label{subsec:off-shell}

The form of the off-shell effective hamiltonian depends on the gauge
chosen for the electroweak gauge fields: by working in the so-called
$R_\xi$ gauge~\cite{Gavela:NPB193}, the simultaneous integration of the $W$
and top fields leaves an effective hamiltonian at the $\mu\simeq M_W$
scale~\cite{inami:lim,deshpande:nazerimonfared} that can be written as
\begin{equation} 
{\cal H}_{\mathrm{eff}} =
\frac{4 G_F}{\sqrt{2}} V^\star_{t\,s} V_{t\,b}\sum_{i=1}^{14} C_i O_i ,
\end{equation}
where the basis of operators $O_i$ invariant under electromagnetic gauge
transformation is
\begin{eqnarray}
O_1 &=& \frac{1}{(4 \pi)^2} \bar{s}_{L}\,\FMslash{D} \FMslash{D}
\FMslash{D}\,b_{L} \nonumber \\
O_2 &=& \frac{(i\,e\,Q_d)}{(4 \pi)^2} \bar{s}_{L}\,\left\{
\FMslash{D},\,F_{\mu \nu}\,\sigma_{\mu \nu} \right\}\,b_{L} \nonumber \\
O_3 &=& \frac{( -\,i\,e\,Q_d)}{(4 \pi)^2}  \bar{s}_{L}\,\gamma_{\nu}\,b_{L}
D_{\mu} F_{\mu \nu} \nonumber \\
O_4 &=&  - \frac{1}{(4 \pi)^{2}} m_{b}\,\bar{s}_{L}\,\FMslash{D}
\FMslash{D}\,b_{R} \nonumber \\
O_5 &=& \frac{( - i\,e\,Q_d) }{(4 \pi)^2} m_{b}\,\bar{s}_{L}\,F_{\mu \nu}
\sigma_{\mu \nu}\,b_{R} \nonumber \\
O_6 &=& \frac{(i\,g_s)}{(4 \pi)^2} \,\bar{s}_{L} \left\{ \FMslash{D},\,T^{A}
G_{\mu \nu}^{A}\,\sigma_{\mu \nu} \right\}\,b_{L} \nonumber \\
O_7 &=& \frac{( - i\,g_s)}{(4 \pi)^2} \,\bar{s}_{L}\,\gamma_{\nu}
T^{A}\,b_{L}\,\left( D_{\mu} G_{\mu \nu} \right)^{A} \nonumber \\
O_8 &=& \frac{( - i\,g_s)}{(4 \pi)^{2}}  m_{b}\,\bar{s}_{L} T^{A}
G^{A}_{\mu \nu} \sigma_{\mu \nu}\,b_{R}\nonumber\\
O_9 &=& \left(\bar{s}_\alpha\gamma_\mu^L
c_\beta\right)\otimes\left(\bar{c}_\beta\gamma_\mu^L b_\alpha\right)
\nonumber\\
O_{10} &=& \left(\bar{s}_\alpha\gamma_\mu^L
c_\alpha\right)\otimes\left(\bar{c}_\beta\gamma_\mu^L
b_\beta\right)\nonumber\\
O_{11/13} &=& \left(\bar{s}_\alpha \gamma_\mu^L b_\alpha\right) \otimes
\sum_{q=u,\dots b} \left(\bar{q}_\beta \gamma_\mu^{L/R}
q_\beta\right)\nonumber\\
O_{12/14} &=& \left(\bar{s}_\alpha \gamma_\mu^L b_\beta\right) \otimes
\sum_{q=u,\dots b}\left(\bar{q}_\beta \gamma_\mu^{L/R} q_\alpha\right)\ .
\label{primabase}
\end{eqnarray}
In the four-fermion operators $O_{9,\cdots,14}$ the Greek indices
refer to the color structure. Note that the operators
$O_{9,11,\dots,14}$ appear only through QCD radiative corrections,
i.e. $C_{9,11,\dots,14}=0$ at the $\mu\simeq M_W$ scale. At the
same scale in the NDR scheme the non-zero coefficients are
($x=m_t^2/M_W^2$) 
\begin{eqnarray}
C_1 &=& {{x\,\left( 1 + 5\,x \right) }\over {2\,{{\left( 1 - x\right) }^3}}}
+ {{{x^2}\,\left( 2 + x \right) \,\log(x)}\over {{{\left( x-1 \right)
}^4}}}\ ,\nonumber\\
C_2 &=& {{x\,\left( -1 - 11\,x + 18\,{x^2} \right) }\over {8\,{{\left( -1 +
x\right)}^3}}} + {{\left( 2 - 5\,x \right) \,x\,\left( -2 + 3\,x \right)
\,\log(x)}\over {4\,{{\left( x-1 \right) }^4}}}\ ,\nonumber\\
C_3 &=& {{-16 + 48\,x - 73\,{x^2} + 35\,{x^3}}\over {12\,{{\left( x-1
\right) }^3}}} + {{\left( 8 - 32\,x + 54\,{x^2} - 30\,{x^3} + 3\,{x^4}
\right) \, \log (x)}\over {6\,{{\left( x-1 \right) }^4}}}\ ,\nonumber\\
C_4 &=& {{x\,\left( 1 + x\right) }\over {2\,{{\left( x-1 \right) }^2}}} +
{{{x^2}\,\log(x)}\over {{{\left( 1 - x \right) }^3}}}\ ,\nonumber\\
C_5 &=& {{\left( 3 - 5\,x \right) \,x}\over {4\,{{\left( x-1 \right)}^2}}}
+ {{x\,\left( -2 + 3\,x \right) \,\log (x)}\over {2\,{{\left( x-1
\right)}^3}}}\ ,\nonumber\\
C_6 &=& {{\left( 4 - x \right)\,x\,\left( -1 + 3\,x\right) }\over
{8\,{{\left( x-1 \right)}^3}}} + {{\left( 2 - 5\,x \right) \,x\,\log
(x)}\over {4\,{{\left( x-1 \right) }^4}}}\ ,\nonumber\\
C_7 &=& {{8 - 42\,x + 35\,{x^2} - 7\,{x^3}}\over {12\,{{\left( x-1 \right)
}^3}}} + {{\left( -4 + 16\,x - 9\,{x^2} \right) \,\log (x)}\over
{6\,{{\left(x-1 \right) }^4}}}\ ,\nonumber\\
C_8 &=& {{\left( -3 + x\right) \,x}\over {4\,{{\left( x-1 \right) }^2}}} +
{{x\,\log(x)}\over {2\,{{\left( x-1 \right) }^3}}}\ ,\nonumber\\
C_{10} &=& 1 . 
\label{eq:offShellCoefficients}
\end{eqnarray}
Note that the operators $O_{1,3,4}$ are proportional to motion
equations. We recall that, given a certain set of renormalized operators
$\N{O}$ which form a complete basis under renormalization, the equations of
motion define combinations which are zero on-shell. These combinations
mix only among themselves in the sense that the AD matrix has the
following block 
form~\cite{kluberg:stern}
\[
\hat{\gamma} = \bordermatrix{
                        & \mathrm{relevant} & \mathrm{motion~equations}\cr
\mathrm{relevant}         & \hat{X}         & \hat{Y}\cr
\mathrm{motion~equations} & 0               & \hat{Z}
}\ .
\]
This AD matrix results from the computation of the renormalization
parts of the 1PI graphs and the relevant operators are chosen
arbitrarily.  
In our first work~\cite{CCRV:plb248} we neglected operators $O_{11}$
and $O_{12}$, which are needed in a complete basis; therefore some
mixings were overlooked.

According to the general strategy exposed in Sec.~\ref{sec:strategy}
we can redefine the effective hamiltonian as
\begin{equation}
{\cal H}_{\mathrm{eff}}\equiv \frac{4\,G_F}{\sqrt{2}} V^\star_{t\,s} V_{t\,b} 
\sum_i \left( C_i^s N[\bar{O}_i^+] + \mbox{`` odd part ''} \right)
\end{equation}
in terms of the ``even and extended'' operators
\begin{equation}
\begin{tabular}{ll}
$\bar{O}^+_1 = \frac{1}{(4 \pi)^2} \bar{s}\,\FMslash{D} \FMslash{D}
\FMslash{D}\,b $ &
$\bar{O}^+_2 = \frac{ (i\,e\,Q_d)}{(4 \pi)^2}\,\bar{s}\,\left\{
\FMslash{D},\,F_{\mu \nu}\,\sigma_{\mu \nu} \right\}\,b $ \\[5pt]
$\bar{O}^+_3 = \frac{ ( -\,i\,e\,Q_d) }{(4 \pi)^2}\,\bar{s}\,\gamma_{\nu}\,b
D_{\mu} F_{\mu \nu} $ & 
$\bar{O}^+_4 =  - \frac{1}{(4 \pi)^{2}} m_{b}\,\bar{s}\,\FMslash{D}
\FMslash{D}\,b
$ \\[5pt]
$\bar{O}^+_5 = \frac{( - i\,e\,Q_d)}{(4 \pi)^2} \, m_{b}\,\bar{s}\,F_{\mu \nu}
\sigma_{\mu \nu}\,b $ &
$\bar{O}^+_6 = \frac{(i\,g_s)}{(4 \pi)^2} \,\bar{s} \left\{ \FMslash{D},\,T^{A}
G_{\mu \nu}^{A}\,\sigma_{\mu \nu} \right\}\,b $ \\[5pt]
$\bar{O}^+_7 = \frac{( - i\,g_s)}{(4 \pi)^2} \,\bar{s}\,\gamma_{\nu}
T^{A}\,b\,\left( D_{\mu} G_{\mu \nu} \right)^{A} $ &
$\bar{O}^+_8 = \frac{( - i\,g_s)}{(4 \pi)^{2}} \, m_{b}\,\bar{s} T^{A}
G^{A}_{\mu\nu} \sigma_{\mu \nu}\,b $ \\[5pt]
$\bar{O}^+_{9,n} = \frac{1}{n!}(\bar{s}_{\alpha} \gb{n}
c_{\beta})\otimes(\bar{c}_{\beta}
\gb{n} b_{\alpha}) $ &
$\bar{O}^+_{10,n} = \frac{1}{n!}(\bar{s}_{\alpha} \gb{n}
c_{\alpha})\otimes(\bar{c}_{\beta}
\gb{n} b_{\beta})  $ \\[5pt]
$\bar{O}^+_{11,n} = \frac{1}{n!}(\bar{s}_{\alpha} \gb{n} 
b_{\alpha})\otimes\sum_q(\bar{q}_{\beta} \gb{n} q_{\beta})$ &
$\bar{O}^+_{12,n} = \frac{1}{n!}(\bar{s}_{\alpha} \gb{n}
b_{\beta})\otimes\sum_q (\bar{q}_{\beta} \gb{n} q_{\alpha})$  \\[5pt]
\end{tabular}\ .
\label{eq:off-shellBasis}
\end{equation}

At leading order the scheme dependence of the coefficients at the 
$\mu\simeq M_W$ scale is irrelevant, for the $b\rightarrow s\gamma$
process: for instance the $C_7$ coefficient depends on the scheme chosen
for the $O_{9,10}$ operators, but it is easy to recognize that on-shell it
gives an $O\left(\alpha_s\right)$ contribution to the $O_{11,13}$
operators, which will be relevant only for a NLO computation.
At the same scale the non-zero coefficients are normalized as follows:
\begin{equation}
\begin{tabular}{lr}
$C^s_i = 2\,C_i\qquad i = 1,\dots 8$ & $C^s_{10,\,1} = C^s_{10,\,3} = C_9$
\end{tabular}\ .
\label{eq:csFromC}
\end{equation}
We shall use the basis~(\ref{eq:off-shellBasis}) in the computation of
the radiative corrections. Note that the index $n$ is arbitrary and
enables us to treat on the same footing relevant and evanescent
operators. 

We use a background field gauge for QCD to avoid the appearance,
off-shell, of non-gauge invariant
operators~\cite{kluberg:stern,collins:book} and ensure that the basis
is closed under renormalization. 

The independence of physical results from  $\xi$ parameter of the
gauge-fixing term 
\[
{\cal L}_{\mathrm{g.f.}} = 
\frac{1}{2\xi}\left(D_\mu\left(G^{\mathrm{classical}}\right) 
G_\mu^{\mathrm{quantum}}\right)^2
\]
will be a useful check of the computation.

\subsection{One loop results}
\label{subsec:oneloop}
\begin{figure}
\begin{center}
\epsfxsize=\textwidth
\leavevmode
\epsfbox{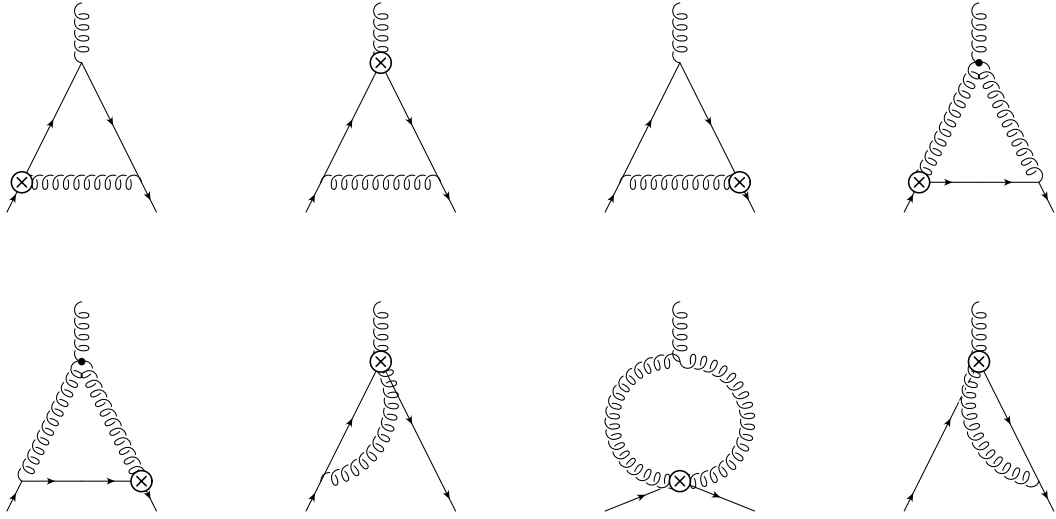}
\end{center}
\caption{Self mixing of operators $\bar{O}^+_{1\dots 8}$.}
\label{fig:mix1-8:1-8}
\end{figure}
\begin{figure}
\begin{center}
\epsfxsize=\textwidth
\leavevmode
\epsfbox{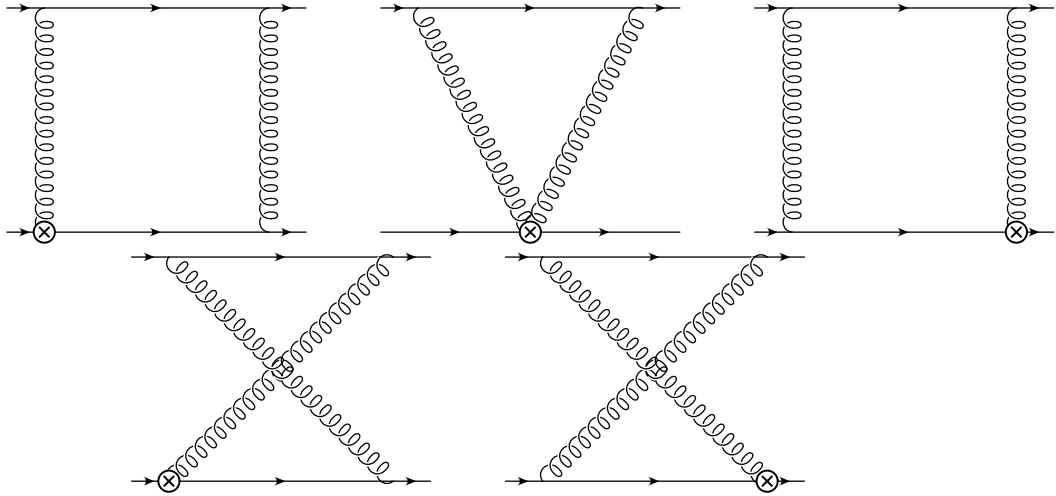}
\end{center}
\caption{Mixing of operators $\bar{O}^+_{1\dots 8}$ with 
$\bar{O}^+_{\left(11,\,n\right)}, \bar{O}^+_{\left(12,\,n\right)}$.}
\label{fig:mix1-8:11-12}
\end{figure}
We list in Fig.~(\ref{fig:mix1-8:1-8}) and in Fig.~(\ref{fig:mix1-8:11-12})
the general structure of the graphs needed to renormalize off-shell the
operators $\bar{O}^+_{1,\dots 8}$: note that the graphs in
Fig.~(\ref{fig:mix1-8:11-12})  are proportional to $\alpha_s^2$, but as
pointed by Misiak~\cite{misiak:plb269} they are required even at leading
order since they give rise to the mixing of the operator $\bar{O}^7$ with
the operators $\bar{O}^+_{\left(11,\,n\right),\,\left(12,\,n\right)}$
in consequence of the motion equations.

The computation of the Feynman graphs in Fig.~(\ref{fig:mix1-8:1-8})
results in the $\left\{1\dots 8\right\}\times \left\{1\dots 8\right\}$
sector of the one-loop anomalous dimension matrix $\hat{\gamma}^{(1)}$,
\begin{equation}
\frac{\alpha_s}{4 \pi}\:
\bordermatrix{
  & 1 & 2 & 3 & 4 & 5 & 6 & 7 & 8 \cr
1 &  {\scriptstyle -2 \,C_F\xi} & 0 & 0 & 
{\scriptstyle 6\,C_F} & 0 & 0 & 0 & 0 \cr
2 & 0 & 0 & 0 & 0 & {\scriptstyle 8\,C_F} & 0 & 0 & 0 \cr
3 & 0 & 0 & 0 & 0 & 0 & 0 & 0 & 0 \cr
4 & 0 & 0 & 0 &  {\scriptstyle 6\,C_F - 2\,C_F\xi} & 0 & 0 & 0 & 0 \cr
5 & 0 & 0 & 0 & 0 & 
{\scriptstyle 8\,C_F }& 0 & 0 & 0 \cr
{\scriptstyle 6} & 
{\scriptstyle - 12\,C_F} & 
{\scriptstyle 4\,C_F} & 0 & 0 & 
{\scriptstyle 4\,C_F} & {\scriptstyle 4\,C_F} & 0 &
{\scriptstyle 4} 
{\scriptstyle ( 3\,C_F - C_A )} \cr
{\scriptstyle 7} & {\scriptstyle 2\,C_F}
 & 0 & {\scriptstyle \frac{8\,C_F}{3}} & 0 & 0 & 0 & 
{\scriptstyle \frac{8}{3}\,C_F + 3\,C_A} & 0 \cr
{\scriptstyle  8} & 0 & 0 & 0 & 
{\scriptstyle - 12\,C_F} & {\scriptstyle  8\,C_F }& 0 & 0 & {\scriptstyle
4 (4\,C_F - C_A )}
}\ ,
\label{eq:mix18:18}
\end{equation}
while the graphs in Fig.~(\ref{fig:mix1-8:11-12}) connect the operator
$\bar{O}^+_7$ to four fermion operators, resulting in 
\begin{equation}
\left(\frac{\alpha_s}{4 \pi}\right)^2\:
\bordermatrix{
  & (11,1) & (11,3) & (12,1) & (12,3) \cr
7 & \frac{-5 + 5\,C_A^2 - 10\,C_A\,C_F}{2} & 
\frac{9\,C_A^2 - 30\,C_A\,C_F + 24\,C_F^2}{2} & 0 & \frac{24\,C_F -
9\,C_A}{2}
}\ .
\label{eq:mix7:11:12}
\end{equation}
\begin{figure}
\begin{center}
\epsfxsize=\textwidth
\leavevmode
\epsfbox{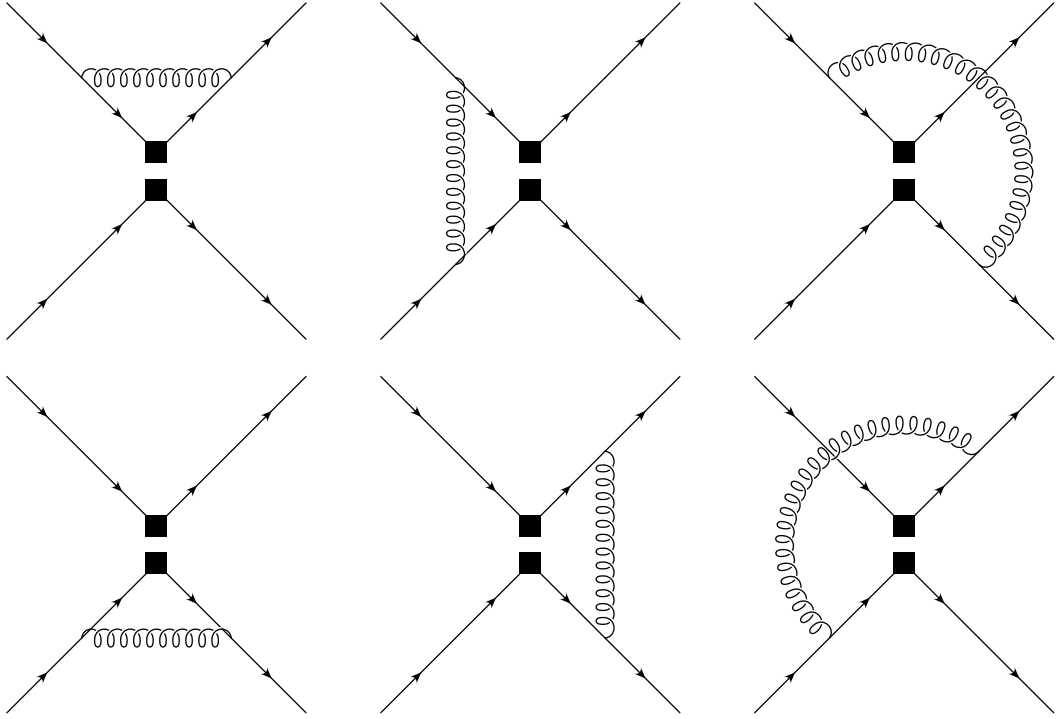}
\end{center}
\caption{\protect{Self mixing of $4$-fermion operators.}}
\protect\label{fig:mix9-12:9-12}
\end{figure}
\begin{figure}
\begin{center}
\epsfxsize=\textwidth
\leavevmode
\epsfbox{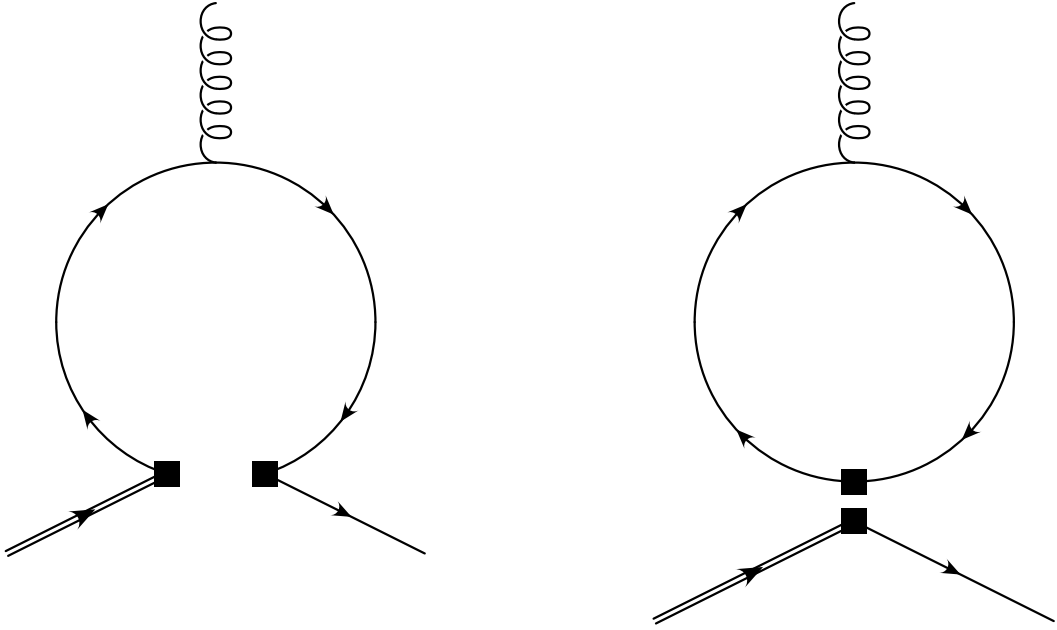}
\end{center}
\caption{\protect{Mixing of $4$-fermion operators with $\bar{O}^+_{1\dots 8}$ 
operators.}}
\protect\label{fig:mix9-12:1-8}
\end{figure}
The renormalization of four fermion operators, off-shell, results from 
the graphs in Fig.~(\ref{fig:mix9-12:9-12}) and in
Fig.~(\ref{fig:mix9-12:1-8}).

The self mixing of four fermion operators at one-loop, in
Fig.~(\ref{fig:mix9-12:9-12}), connects Dirac structures with $\Delta
n = {+2,\,0,\,-2}$
\begin{equation}
\frac{\alpha_s}{4 \pi}\:\bordermatrix{
     &  (9,n-2) & (9,n) & (9,n+2) \cr
(9,n) & 
{{\left( 6 - n \right) \,\left(n - 5\right) }\over {C_A}} & 
{\scriptstyle 2\,C_F\,\left( \left( 1 - n \right) \,\left(n - 3\right) - 
\xi  \right)} &
{{-\left( \left( 1 + n \right) \,\left( 2 + n \right)  \right) }\over 
    {C_A}} \cr
(10,n) & 
{{\left(n - 6\right) \,\left(n - 5\right) }\over 2} &
{\scriptstyle 3\,\left( -2 + 4\,n - {n^2} \right) } &
{{\left( 1 + n \right) \,\left( 2 + n \right) }\over 2}\cr
 }\ ,
\end{equation}
\begin{equation}
\frac{\alpha_s}{4 \pi}\:\bordermatrix{
     &  (10,n-2) & (10,n) & (10,n+2) \cr
(9,n) &
{\scriptstyle \left(n - 6\right) \,\left(n - 5\right) } &
0 &
{\scriptstyle \left( 1 + n \right) \,\left( 2 + n \right) } \cr
(10,n) &
{{\left(4\,C_F -\,C_A\right) \,\left(n - 6\right) \,
      \left(n - 5\right) }\over 2} &
{\scriptstyle \left(3\,C_A - 4\,C_F\right)\,\left( 2 - 4\,n + {n^2} \right) - 
      4\,C_F\,\left( 1 + \xi  \right) } &
{{\left(4\,C_F -\,C_A\right) \,\left( 1 + n \right) \,
      \left( 2 + n \right) }\over 2}
\cr
 }\ .
\end{equation}
The same mixings occur in the $\bar{O}^+_{(11,n),(12,n)}$ sector.

The penguin graphs in Fig.~(\ref{fig:mix9-12:1-8}) give rise
to the mixing of the 4-fermion operators with $\bar{O}^+_{3,\,7}$ at
the $\alpha_s^0$ order
\begin{equation}
\bordermatrix{
  & 3 & 7\cr
(9,n)  & -\frac{2\left(n - 2\right)}{3 n!} \left(-1\right)^n c_n(0) & 
\frac{(n-2)}{3 n!} (-1)^n c_n(0)\cr
(10,n) & -C_A\frac{2\left(n-2\right)}{3 n!} (-1)^n c_n(0) & 0\cr
(11,n) & C_A\frac{8}{3}\bar{n}_f\delta_{n,1} + 
\frac{2\left(n-2\right)}{3 n!} (-1)^n c_n(0) & \frac{2\left(n-2\right)}{3
n!} (-1)^n c_n(0) \cr
(12,n) & \frac{8}{3}\bar{n}_f\delta_{n,1} +
C_A \frac{2\left(n-2\right)}{3 n!} (-1)^n c_n(0) & \frac{8}{3} n_f
\delta_{n,1}\cr
}\ .
\end{equation}
The meaning of the symbol $c_n$ is explained in 
App.~\ref{app:dirac} together with the other definitions and useful
formulas. We use the symbols $n_f=u+d,\,\bar{n}_f = 
d-2\,u$, with $u$ and $d$ 
being the number of up and down  quark species active.

It is to be noted that at leading order only the $n=\{1,\,3,\,5\}$ values
are needed, because the effective hamiltonian starts with $n=\{1,\,3\}$ and 
at one loop only the $n=5$ evanescent arises.
We find convenient to give also the results restricted to this set of
$n$ values 
\begin{equation}
\frac{\alpha_s}{4 \pi}\:
(\bordermatrix{
     &  (9,1) & (9,3) & (9,5) & (10,1) & (10,3) & (10,5) \cr
(9,1) & 0 & - \frac{6}{C_A} & 0 & 0 & 6 & 0 \cr
(9,3) &  - \frac{6}{C_A} &  0 & -\frac{20}{C_A} & 6 & 0 & 20\cr
(10,1) & 3 & 3 & 0 & - \frac{3}{C_A} - 6\,C_F & - \frac{3}{C_A} + 6\,C_F & 0
\cr
(10,3) & 3 & 3 & 10 & - \frac{3}{C_A} + 6\,C_F & - \frac{3}{C_A} - 6\,C_F & 
20\,C_F - \frac{10}{C_A} \cr
 }\ ,
\label{eq:mix910:910}
\end{equation}
\begin{equation}
\bordermatrix{
         &   3   &  7\cr
(9,\,1)  &  -\frac{8}{3} & \frac{4}{3}\cr
(9,\,3)  & - \frac{8}{3} & \frac{4}{3}\cr
(10,\,1) & -\frac{8\,C_A}{3} & 0\cr
(10,\,3) & - \frac{8\,C_A}{3} & 0\cr
(11,\,1) & \frac{8}{3}\left(1 + C_A\,\bar{n}_f\right) & \frac{8}{3}\cr
(11,\,3) & \frac{8}{3} & \frac{8}{3}\cr
(12,\,1) & \frac{8}{3}\left(C_A + \,\bar{n}_f\right) & \frac{8}{3} n_f\cr
(12,\,3) & \frac{8}{3}\,C_A & 0
}\ .
\end{equation}
It is well known that renormalized operators proportional to motion equations
mix only between themselves and that their anomalous dimension matrix is not
gauge independent. In fact, we can observe that operator $\N{\bar{O}^+_1}$, 
proportional to the $s$ motion equation, mixes only with itself and 
operator $\N{\bar{O}^s_4}$: this one mixes with itself only, and both
operators have anomalous dimension matrix depending on
$\xi$~\cite{kluberg:stern}. 
The difference $\N{\bar{O}^+_2 -\bar{O}^+_5}$ is 
proportional to a combination of the $s$ and $b$ motion equations and 
does not evolve at one-loop. Analogously the operator $\N{\bar{O}^+_6
- \bar{O}^+_8}$ 
mixes with itself and with $\N{\bar{O}^+_1},\,\N{\bar{O}^+_2 -
\bar{O}^+_5},\,\N{\bar{O}^+_4}$. 
Finally the combination
\[
\N{\bar{O}^+_7 + \frac{\alpha_s}{4\pi}\left(\frac{1}{2\,C_A}
\bar{O}^+_{(11,\,1)} - \frac{1}{2} \bar{O}^+_{(12,\,1)}\right)}
\]
is proportional to the equation of motion of the
gluon. It is worth noting that the elimination of the operator
$\N{\bar{O}^+_7}$ in favor of the four fermion operators introduces a factor 
$\frac{1}{\alpha_s}$ which combines with the $\alpha_s^2$ in 
Eq.~(\ref{eq:mix7:11:12}) to give a result relevant for the leading order 
computation~\cite{misiak:plb269}.

\subsection{Two loop results}
\label{subsec:twoloop}

\begin{figure}
\begin{center}
\epsfxsize=0.9\textwidth
\leavevmode
\epsfbox{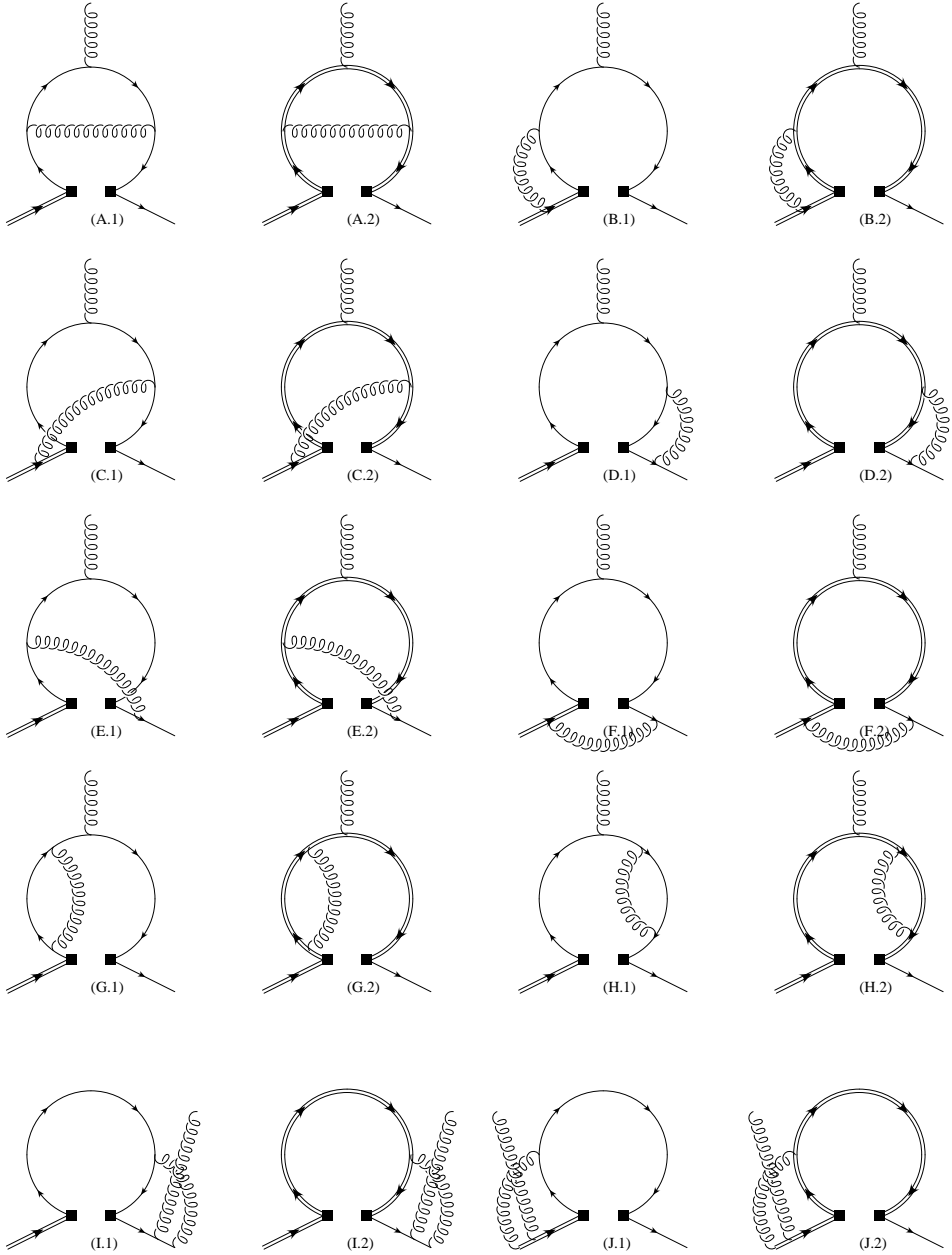}
\end{center}
\caption{``Open'' graphs contributing to the renormalization of
$\bar{O}^+_{\left(9-12,\,n\right)}$.}
\label{fig:mix21}
\end{figure}
\begin{figure}
\begin{center}
\epsfxsize=0.9\textwidth
\leavevmode
\epsfbox{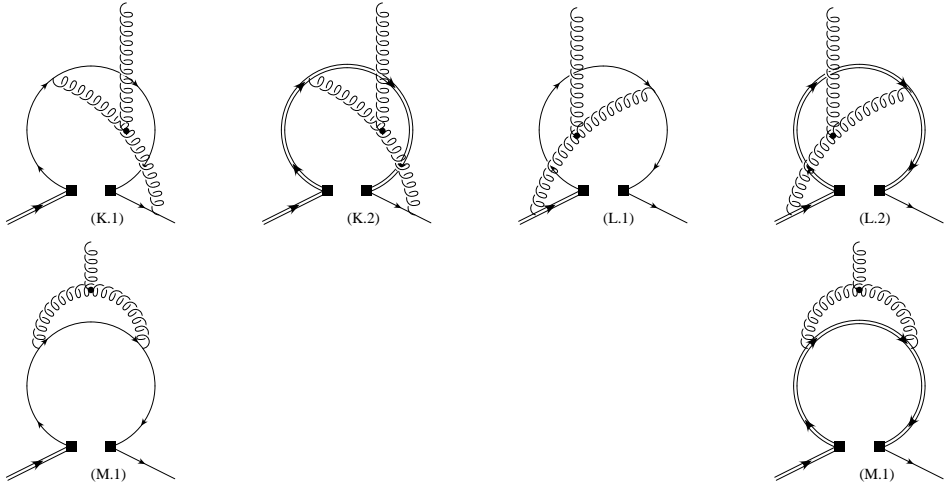}
\end{center}
\caption{``Open'' graphs for the renormalization of
$\bar{O}^+_{\left(9-12,\,n\right)}$. Non abelian couplings.}
\label{fig:mix21a}
\end{figure}
\begin{figure}
\begin{center}
\epsfxsize=0.9\textwidth
\leavevmode
\epsfbox{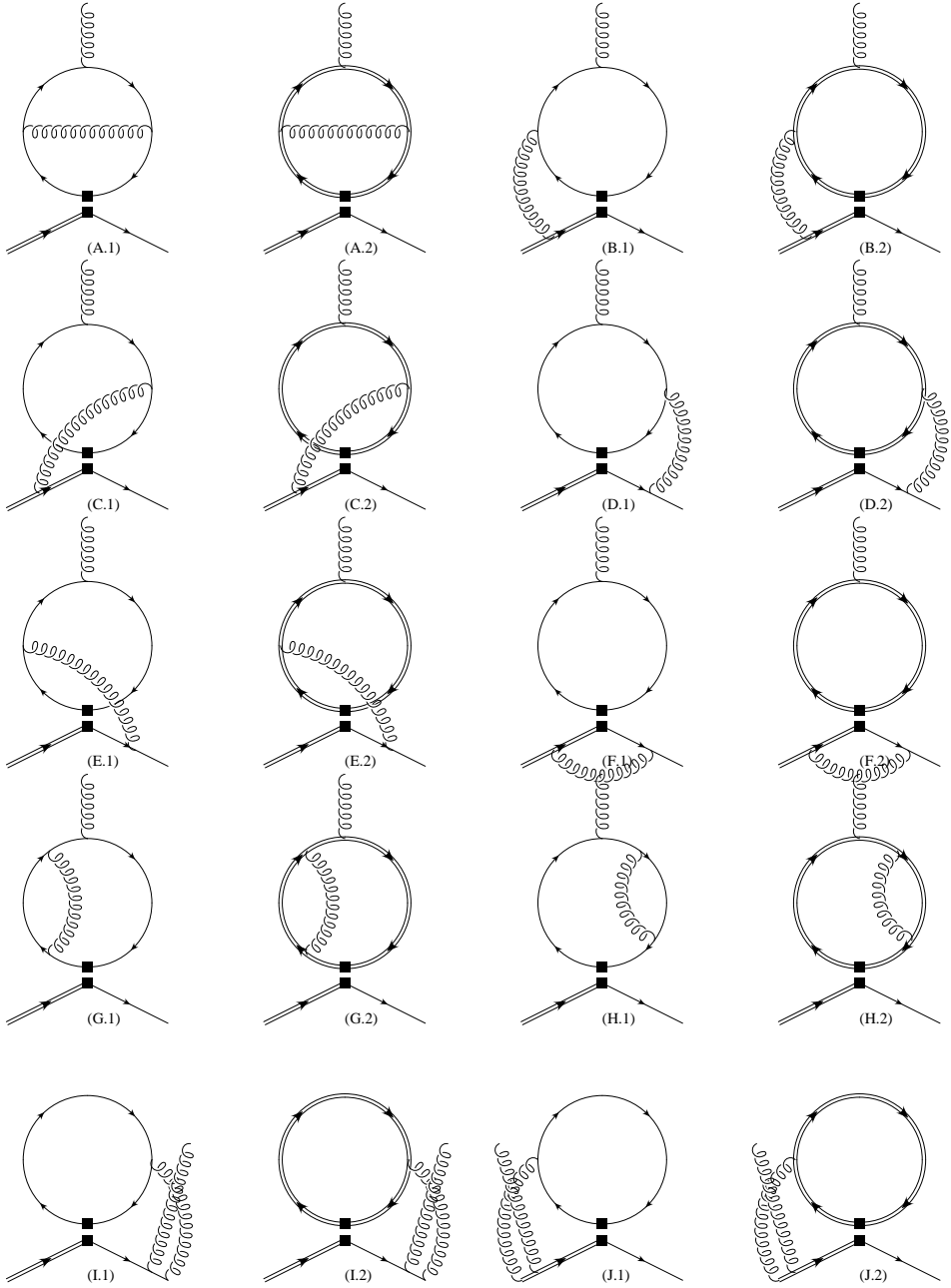}
\end{center}
\caption{`Closed'' graphs contributing to the renormalization of
$\bar{O}^+_{\left(11-12,\,n\right)}$.}
\label{fig:mix22}
\end{figure}
\begin{figure}
\begin{center}
\epsfxsize=0.9\textwidth
\leavevmode
\epsfbox{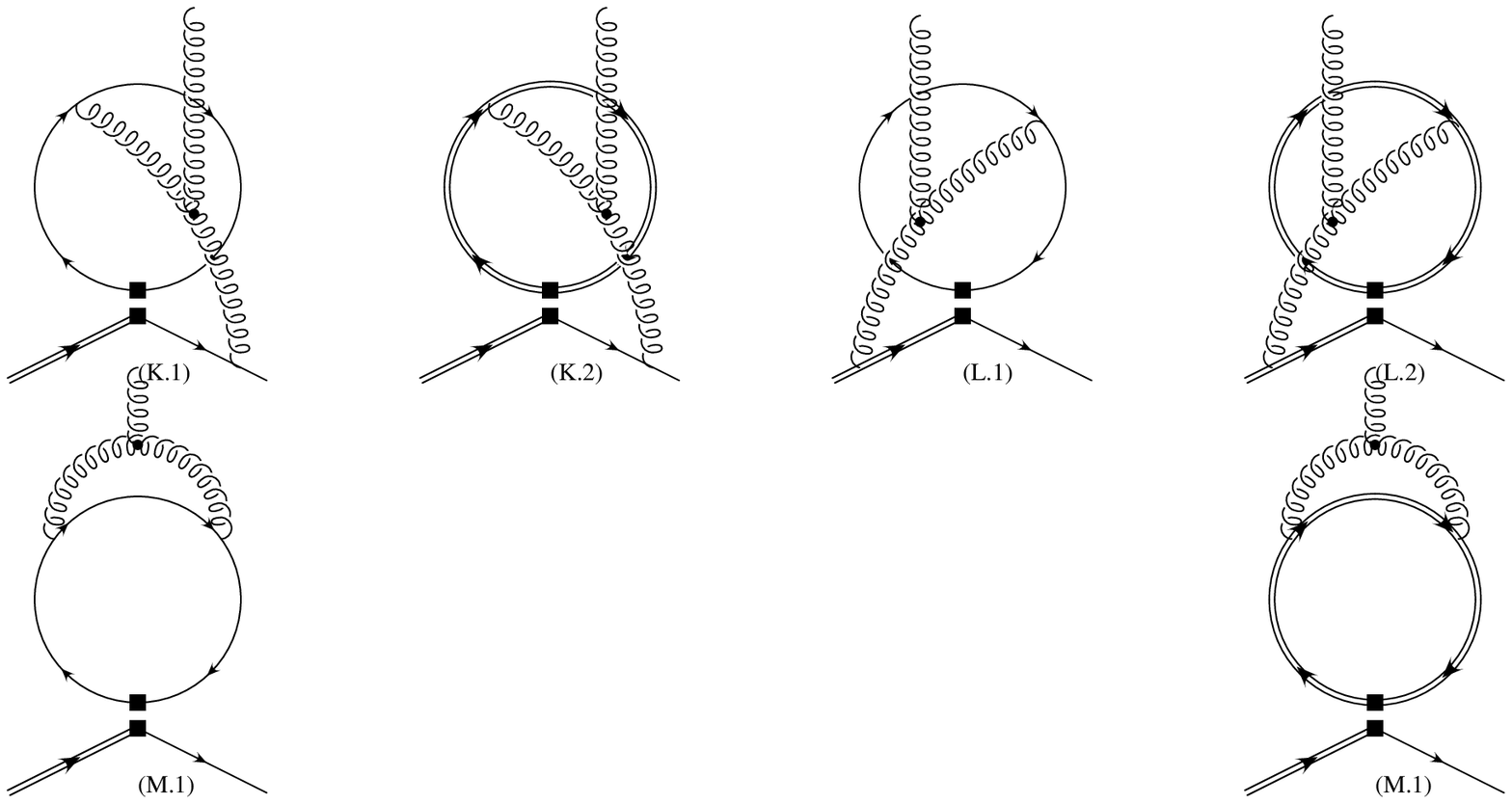}
\end{center}
\caption{``Closed'' graphs for the renormalization of
$\bar{O}^+_{\left(11-12,\,n\right)}$. Non abelian couplings.}
\protect\label{fig:mix22a}
\end{figure}

The two-loop mixing of the four fermion operators with operators 
$\bar{O}^+_{1,\dots 8}$ can be obtained from the computation of the Feynman 
diagrams in Figs.~(\ref{fig:mix21},~\ref{fig:mix21a})
and in Figs.~(\ref{fig:mix22},~\ref{fig:mix22a}):
the ``closed'' loop graphs are relevant only for the renormalization of
operators $\bar{O}^+_{\left(\left(11-12\right),\,n\right)}$.

The massive $b$ quark is represented by the heavy-faced lines.
The $b$ propagator is expanded in series 
of $m_b$ up to the first order and the resulting massless graphs are 
regularized in the infrared region by the flow of the external momenta: the
diagrams have to be evaluated with zero and one mass insertions. The loop
integrals have been computed with the help of the algorithms developed by
Chetyrkin et al.~\cite{chetyrkin:tkachov:npb192}, implemented in the {\it
Mathematica}~\cite{math} symbolic manipulation language.
\begin{table}
\caption{Two loop ADM entries for
$\bar{O}^+_{\left(9,\,n\right)},\,\bar{O}^+_{\left(10,\,n\right)}$, resulting
from the mixing with $\bar{O}^+_1,\,\bar{O}^+_8$}
\label{tab:adm9:10}
\begin{tabular}{ll}\hline
  $\gamma$ entry & 
  \\ 
  ${(9, n), 1}$ 
  & 
  $\left(-1\right)^n \frac{C_F}{3 n!}\left(\left(1 - n\right) c_n(0) + 
  \left(2 - n\right)  c_n(1)\right)$ 
  \\ 
  ${(9, n), 2}$ & $\frac{2\,C_F}{n!}\left[ -f_{n,3}\left(0\right) + 
  \left(-1\right)^n \frac{\left(2-n\right)}{9} c_n\left(0\right)\right]$ 
  \\ 
  ${(9, n), 3}$ 
  & 
  $ \frac{4\,C_F}{n!}\left[- f_{n,3}(0) + \frac{n (n-1)(-1)^n}{18} 
  \left(1908 - 1124 n + 213 n^2 - 13 n^3\right) c_{n-2}(0)\right.$ 
  \\ 
  & 
  $\left.\qquad + \frac{(-1)^n}{54}\left(160 - 290 n + 177 n^2 - 33
  n^3\right) c_n(0) + \frac{2 (n-2) (-1)^n}{9}\left(4 - 12 n + 3 n^2\right)
  c_n(1)\right]$
  \\ 
  ${(9, n), 4}$ 
  &
  $\frac{C_F\,\left(2-n\right)\left(-1\right)^n}{2 n!} c_n(0)$ 
  \\ 
  ${(9, n), 5}$ 
  & 
  $\frac{2\,C_F\,\left(n-2\right)}{3 n!}\left(2 n^2 - 8 n + 3\right) 
  \left(-1\right)^n c_n(0)$ 
  \\ 
  ${(9, n), 6}$ 
  & 
  $\frac{4\,C_F - C_A}{4 n!} f_{n,3}(0) + \left(-1\right)^n 
  \frac{\left(8\,C_F - 9\,C_A\right) \left(2 - n\right)}{36 n!} c_n(0)$
  \\ 
  ${(9, n), 7}$ 
  & 
  ${{\left( 4\,C_F - C_A \right) \,f_{n,3}(0)}\over{2 n!}} + 
  {{{{\left( -1 \right) }^n}\,\left( C_A - 2\,C_F \right) \,
  n\,\left( n - 1 \right)} \over {18 n!}} \,
  \left( 1908 - 1124\,n + 213\,{n^2} - 13\,{n^3} \right) \, c_{n-2}(0)$
  \\ 
  & 
  $+ {{\left( -1 \right)^n} \over {108 n!}} \,\left( 3\,C_A\,
  \left(4-n\right) \,
  \left( 3 - 46\,n + 24\,{n^2} \right) - 
  4\,C_F\,\left(58 - 230\,n + 177\,{n^2} - 33\,{n^3} \right) \right)
  \,c_n(0)$
  \\ 
  & 
  $+{{\left( -1 \right)^n} \over {18 n!}} \,\left( n-2 \right) \,
  \left( 4\,C_F\,\left( -11 + 24\,n - 6\,{n^2} \right) + 3\,C_A\,
  \left(-3 - 8\,n + 2\,{n^2} \right) \right) \, c_n(1)$ 
  \\ 
  ${(9, n), 8}$ 
  & 
  ${{{{\left( -1 \right) }^n}\,\left( n-2 \right)} \over {24 n!}} \,
  \left( 8\,C_F\,\left( -3 + 8\,n - 2\,{n^2} \right) + C_A\,\left( 1 -
  16\,n + 4\,{n^2} \right) \right) \,c_n(0)$
  \\ 
  \hline
  ${(10, n), 1\,(2)}$ 
  &
  0 
  \\ 
  ${(10, n), 3}$ 
  & 
  $\frac{C_A\,C_F \left(-1\right)^n}{9 n!} 
  \left[n\,\left( n - 1\right)\,\left( 1908-1124\,n + 213\,{n^2} -
  13\,{n^3} \right) \,
  c_{n-2}(0)\right.$ 
  \\ 
  &
  $\left. \qquad + \left( 108 - 148\,n + 83\,{n^2} - 13\,{n^3} \right)
  \,c_n(0)\right]$ 
  \\ 
  ${(10, n), 4\,(5)}$ 
  & 
  $0$ 
  \\ 
  ${(10, n), 6}$ 
  & 
  $\frac{1}{2 n!} f_{n,3}(0)$ 
  \\ 
  ${(10, n), 7}$ 
  &
  $\frac{1}{n!} f_{n,3}(0) + {{{{\left( -1 \right) }^n}\,n\,
  \left(n-1\right)} \over {36 n!}}\,
  \left( -1908 + 1124\,n - 213\,{n^2} + 13\,{n^3} \right) \, c_{n-2}(0)$ 
  \\ 
  & 
  $+{{\left( -1 \right)^n} \over {36 n!}}\,
  \left( 24 + 32\,n - 35\,{n^2} + 9\,{n^3} \right) \,
  c_n(0) + {{{{\left( -1 \right) }^n}\,\left( n - 2\right)} \over {3 n!}} \,
  \left( -3 + 8\,n - 2\,{n^2} \right) \,c_n(1)$
  \\ 
  ${(10, n), 8}$ 
  &
  ${{{{\left( -1 \right) }^n}\,\left( n - 2 \right)} \over {6 n!}}
  \,\left( -3 + 8\,n - 2\,{n^2} \right) \,c_n(0)$ \\ 
\hline
\end{tabular}
\end{table}
\begin{table}
\caption{Two loop ADM entries for $\bar{O}^+_{\left(11,\,n\right)}$,
resulting from the mixing with $\bar{O}^+_1,\,\bar{O}^+_8$}
\label{tab:adm11}
\begin{tabular}{ll}\hline
  $\gamma$ entry 
  & 
  \\ 
  ${(11, n), 1}$ 
  &
  $\frac{2\,C_F\,\left(-1\right)^n}{3\,n!}\left[\left(1-n\right)
  c_n(0) + \left(2-n\right) c_n(1)\right]$ 
  \\ 
  ${(11, n), 2}$ 
  &
  $\frac{2\,C_F}{n!}\left[f_{n,\,3}(0) + \frac{2}{9}\left(2-n\right)
  \left(-1\right)^n c_n(0)\right]$ 
  \\ 
  ${(11, n), 3}$ 
  &
  $\frac{C_F}{n!}\left[8\,C_A\,\bar{n}_f \delta_{n,\,1} + 
  4\,f_{n,\,3}(0)\right.$
  \\ 
  & 
  $\qquad +\frac{2}{9}\,\left(-1\right)^n\,\left(n - 1\right) \,n\, 
  \left( -1908 + 1124\,n - 213\,{n^2} + 13\,{n^3} \right) \,c_{n-1}(0)$ 
  \\ 
  & 
  $\qquad +\frac{2}{27}\,\left(-1\right)^n\,
  \left(-58 + 230\,n - 177\,{n^2} + 33\,{n^3} \right) \,c_n(0)$ 
  \\ 
  & 
  $\qquad\left. +\frac{4}{9}\,(-1)^n\,\left( n-2 \right) \,
  \left( -11 + 24\,n - 6\,{n^2} \right) \,c_n(1)\right]$ 
  \\ 
  ${(11, n), 4}$ 
  &
  $\frac{C_F}{n!} 
  \left[\frac{1}{3} {\left( 39 - 68\,n + 20\,{n^2} + 
    3\,\left(2-n\right)\left(-1\right)^n\right) \,c_n(0)} + 2\,
  \left(n-3 \right) \,\left( n - 1\right) \,c_n(1)\right]$ 
  \\ 
  ${(11, n), 5}$ 
  &
  $ \frac{C_F}{3 n!}\left[2\,\left( n-6 \right) \,
  {{\left( n-5 \right)}^2}\, \left( n-3 \right) \,
  \left( n - 1 \right)\,n\,c_{n-2}(1)\right.$ 
  \\ 
  & 
  $ \qquad -\frac{1}{3}{\left( 45 + 100\,n - 170\,{n^2} + 56\,{n^3} -
  4\,{n^4} +
  2\,\left(n-2\right)\,\left(2\,n^2- 8\,n +
  3\right)\left(-1\right)^n\right) \,c_n(0)} $
  \\ 
  & 
  $\left. \qquad + 2\,\left( n-3 \right) \,\left( n - 1 \right) \, 
  \left( -5 + 3\,n - {n^2} \right) \,c_n(1)\right]$ 
  \\ 
  ${(11, n), 6}$ 
  & 
  $\frac{1}{n!}\left[12\,n_f\delta_{n,3} + \frac{1}{2} 
  \left(4\,C_F - C_A\right) f_{n,3}(0)\right.$ 
  \\ 
  & 
  $\qquad\left.+ \frac{\left(-1\right)^n}{18}\left(9\,C_A-8\,C_F\right)
  \left(n-2\right) c_n(0)\right]$
  \\ 
  ${(11, n), 7}$ 
  &
  $\frac{1}{n!}\left[88\,n_f \delta_{n,3} + \left(4\,C_F - C_A\right)
  f_{n,3}(0)\right.$ 
  \\ 
  & 
  $\qquad + \frac{1}{9}\left(-1\right)^n\,\left(2\,C_F - C_A\right)
  \,\left(n-1\right) \,n\,\left( -1908 + 1124\,n - 213\,{n^2} +
  13\,{n^3} \right) \,c_{n-2}(0)$ 
  \\ 
  & 
  $\qquad + \frac{1}{54}\left(-1\right)^n \left( 3\,C_A\,\left( n-4 \right)
  \,\left( -3 + 46\,n - 24\,{n^2} \right) \right. +$ 
  \\
  &
  $\qquad +\left. 4\,\,C_F\,\left( -58 + 230\,n - 177\,{n^2} +
  33\,{n^3} \right)\right)\,c_n(0)$
  \\ 
  & 
  $\left.\qquad +\frac{1}{9}\left(-1\right)^n\left( 2 - n \right) \,
  \left( 3\,C_A\,\left( 3 + 8\,n - 2\,{n^2} \right) + 4\,C_F\,
  \left( 11 - 24\,n + 6\,{n^2} \right) \right) \, c_n(1)\right]$
  \\ 
  ${(11, n), 8}$ 
  &
  $\frac{1}{n!}\left[\frac{1}{3}{\left( -C_A + 2\,C_F \right) \,
    \left(n-6 \right) \, {{\left( n-5 \right) }^2}\,\left( n-3 \right) \,
    \left( n - 1 \right) \,n\,c_{n-2}(1)}\right.$ 
  \\ 
  & 
  $\qquad + \frac{1}{18}\left( C_A\,\left( 99 + 24\,n - 125\,{n^2} +  
  40\,{n^3}- 2\,{n^4} \right) + \right.$ 
  \\ 
  &
  $\qquad + \left. 2\,C_F\,\left(-45 - 100\,n + 170\,{n^2} - 56\,
  {n^3} + 4\,{n^4} \right) \right) \,c_n(0)$ 
  \\ 
  & 
  $\qquad + \frac{1}{3}{\left( n-3 \right) \,\left(n-1 \right) \, 
    \left( C_A\,n + 2\,C_F\, \left( -5 + 3\,n - {n^2}\right) \right) \,
    c_n(1)} + 24\,n_f\,\delta_{n,3}$ 
  \\ 
  & 
  $\qquad \left. +\frac{1}{12}\left(-1\right)^n\left(n - 2\right) \,
  \left(8\,C_F\,\left( -3 + 8\,n - 2\,{n^2} \right) + C_A\,
  \left( 1 - 16\,n +4\,{n^2} \right) \right) \,c_n(0) \right]$\\
\hline
\end{tabular}
\end{table}
\begin{table}
\caption{Two loop ADM entries for $\bar{O}^+_{\left(12,\,n\right)}$,
resulting from the mixing with $\bar{O}^+_1,\,\bar{O}^+_8$}
\label{tab:adm12}
\begin{tabular}{ll}\hline
  $\gamma$ entry 
  & 
  \\
  ${(12, n), 1}$
  &
  $-\frac{4}{3}\,C_F n_f \delta_{n, 1}$
  \\
  ${(12, n), 2}$
  &
  $-\frac{16}{9}\,C_F n_f \delta_{n, 1} + 4\,C_F\bar{n}_f \delta_{n, 3}$
  \\
  ${(12, n), 3}$
  &
   $-\frac{8}{81}\,C_F\left(102 n_f - 81 \bar{n}_f\right)
    \delta_{n,1} + \frac{88}{3}\,C_F \bar{n}_f \delta_{n, 3}$
  \\
  &
  $\qquad + \frac{C_A\,C_F\,\left(-1\right)^n}{9\,n!}\left[ \left( n -
1\right)\,n\,\left( -1908 + 1124\,n - 213\,{n^2} + 13\,{n^3} \right)
\,c_{n-2}(0)\right.$
  \\
  &
  $\qquad\qquad\left. + \left( -108 + 148\,n - 83\,{n^2} + 13\,{n^3}
  \right) \,c_n(0)\right]$\\ ${(12, n), 4}$ & $- 4\,C_F n_f \delta_{n,1}$
  \\
  ${(12, n), 5}$
  &
  $\frac{C_A\,C_F}{3\,n!}\,\left[\left( n-6 \right)\,{{\left( n-5 \right)
}^2}\, \left(n-3\right)\,\left( n - 1 \right)\,n\,c_{n-2}(1)\right.$
  \\ 
  &
  $\qquad + \frac{1}{3}\left( 45 - 80\,n + 56\,{n^2} -
   12\,{n^3}\right)\,c_n(0)$
  \\ 
  &
  $\left. \qquad +\left( n-3 \right) \, \left( n -1 \right) \,\left( -2 -
   5\,n + {n^2} \right) \, c_n(1)\right] +8\,C_F\,\bar{n}_f\,\delta(n,3)$
  \\
  ${(12, n), 6}$
  &
  $\frac{1}{n!}\,f_{n,3}(0) + \frac{2}{9}\,\left(9\,C_A - 8\,C_F\right)
  \,n_f\, \delta_{n,1} + \left(4\,C_F - C_A\right) \,n_f \,\delta_{n,3} $
  \\
  ${(12, n), 7}$
  &
  $\frac{28}{27} \left(9\,C_A - 2\,C_F\right)n_f\delta_{n, 1} + \frac{22}{3}
   \left(4\,C_F - C_A\right) n_f\delta_{n,3} + \frac{2}{n!} f_{n,3}(0)$
  \\
  &
  $\frac{\left(-1\right)^n}{3 n!}\left[\frac{1}{6}\left( n-1
  \right)\,n\,\left( -1908 + 1124\,n - 213\,{n^2} + 13\,{n^3} \right)
  \,c_{n-2}(0)\right.$
  \\
  &
  $\qquad + \frac{1}{6}\left( 24 + 32\,n - 35\,{n^2} +
   9\,{n^3}\right)\,c_n(0)$
  \\
  &
  $\qquad + 2\,\left( 2 - n\right) \,\left( 3 - 8\,n + 2\,{n^2} \right)
   \,c_n(1)$ \\
  ${(12, n), 8}$
  &
  $\frac{-5\,C_A}{3} n_f \delta_{n,1} + 2\,\left(4\,C_F - C_A\right)\,n_f
\delta_{n,3}$
  \\
  &
  $+\frac{1}{3\,n!}\left[\frac{1}{2}\left(n-6\right)\,
  {{\left(n-5\right)}^2}\,\left(n-3\right)\,\left(n-1\right)\,
  n\,c_{n-2}(1)\right.$
  \\
  &
  $\quad + \frac{1}{3}\left(73\,{n^2} -18 - 62\,n - 22\,{n^3} + 2\,{n^4}
   \left(-1\right)^n\,\left(18 - 57\,n + 36\,n^2 -6\,n^3\right)\right)
   \,c_n(0)$
  \\
  &
  $\left. \quad +\frac{1}{2}\left( n - 3 \right) \,\left( n - 1\right)
   \,\left( -4 + 11\,n - 3\,{n^2} \right) \,c_n(1)\right]$
  \\
\hline
\end{tabular}
\end{table}
As in the case of the one loop computation, the results can be given for 
arbitrary $n$ and are listed in 
Tabb.~(\ref{tab:adm9:10},~\ref{tab:adm11},~\ref{tab:adm12}). We 
refer to App.~\ref{app:dirac} for symbols and definitions; let us just
note that the symbol $f_{n\,j}$ results from traces involving elements
of the $\{\gb{n}\}$ basis.

We stress that having the results for arbitrary $n$ will be useful for
the NLO computation.

\subsection{Reduction of evanescent operators}
\label{subsec:evanescents}
\begin{figure}
\begin{center}
\epsfxsize=\textwidth
\leavevmode
\epsfbox{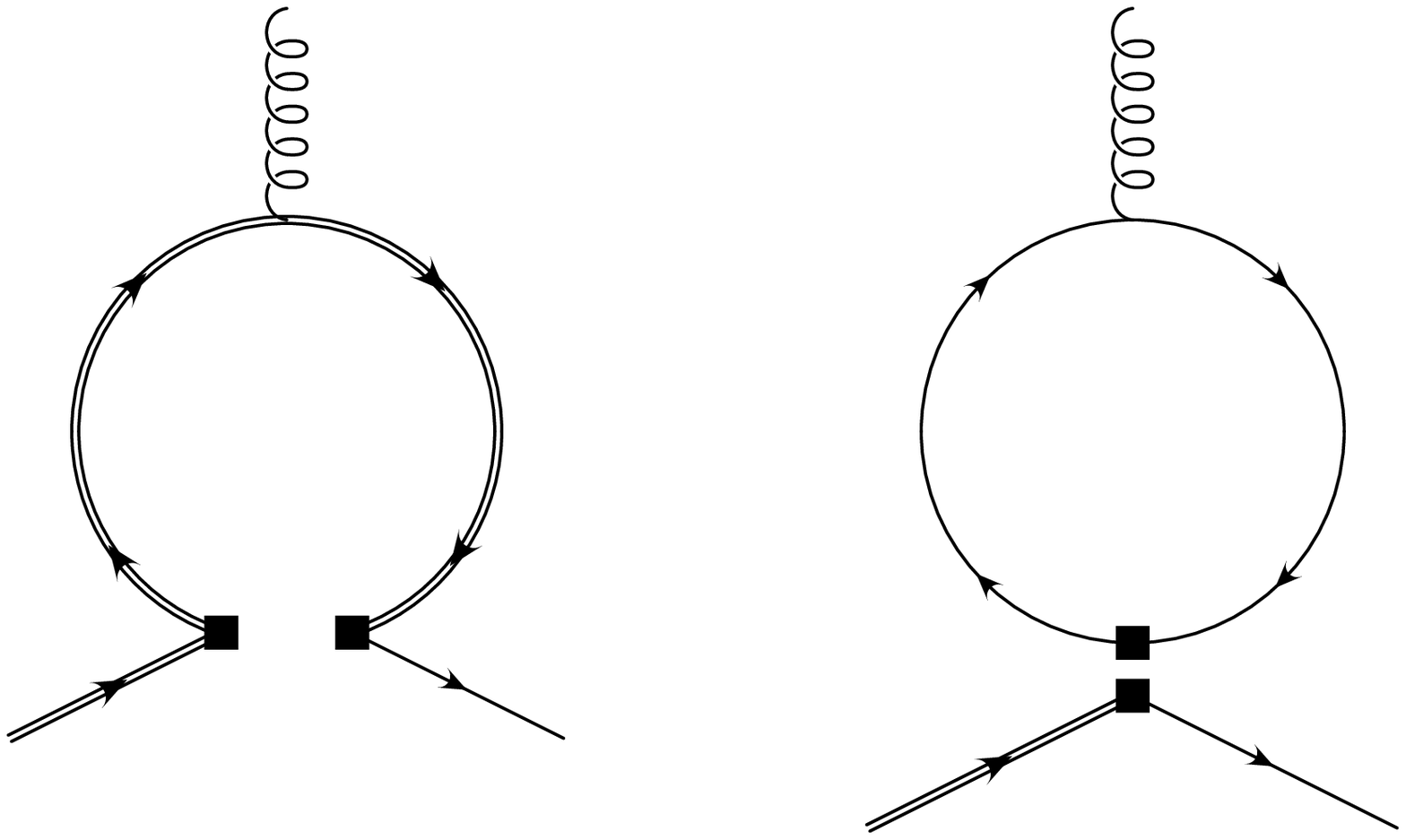}
\end{center}
\caption{Diagrams relevant at LO for the reduction of evanescent
operators.}
\protect\label{fig:redev}
\end{figure}
Before giving the results for $n=1,\,3$, that are needed in this 
leading logarithmic computation, we perform the reduction of the
evanescent operators with $n=5$. The scheme we follow is detailed in 
App.~\ref{subsec:redev}, and requires the computation of the graphs in 
Fig.~(\ref{fig:redev}), which allows to express the insertion 
of the evanescent operators in the Green functions in terms of the
insertion of relevant operators, with coefficients addressed by the
matrix $\hat{r}$
\begin{equation}
\hat{r} = 
\bordermatrix{
       & 3 & 5 & 7 & 8 \cr
(9,5)  & \frac{2}{5} & 0 & -\frac{1}{5} & 0 \cr
(10,5) & \frac{2}{5}\,C_A & 0 & 0 & 0 \cr
(11,5) & -\frac{2}{5} & \frac{8}{15} & -\frac{2}{5} & \frac{8}{15} \cr
(12,5) & -\frac{2}{5}\,C_A & \frac{8}{15}\,C_A & 0 & 0\cr
} .
\label{eq:redev}
\end{equation}
For the present computation only columns $5,\,8$ are relevant, while
the columns $3,\,7$ contribute through equations of motion to four
fermion operators, 
\begin{equation}
\hat{r}_{\mathrm{on-shell}} = \frac{\alpha_s}{4\pi} \bordermatrix{
       & (11,1) & (12,1) \cr
(9,5)  & \frac{1}{10\,C_A} & - \frac{1}{10} \cr 
(10,5) & 0 & 0 \cr
(11,5) & \frac{1}{5\,C_A} & - \frac{1}{5} \cr
(12,5) & 0 & 0\cr
}\ ,
\end{equation}
and are relevant only for the NLO computation.

By combining the reduction coefficients in Eq.~(\ref{eq:redev}), the one 
loop anomalous dimension matrix in Eq.~(\ref{eq:mix910:910}) and the 
results of the two loop computation listed in 
Tabb.~(\ref{tab:adm9:10},~\ref{tab:adm11},~\ref{tab:adm12}), we are able to
give the anomalous dimension matrix relevant at leading order, where a factor
$\frac{\alpha_s}{4\,\pi}$ is understood.
\begin{equation}
\bordermatrix{
         & 1 & 2 & 3 & 4\cr
(9,\,1) & {{2\,C_F}\over 3} & {{-44\,C_F}\over 9} & {{-88\,C_F}\over {27}}
& -2\,C_F \cr
(9,\,3) & {{2\,C_F}\over 9} & {{-44\,C_F}\over 9} & 8\,C_A -
{{1688\,C_F}\over {27}} & -2\,C_F \cr
(10,\,1) & 0 & 0 & {{-40\,C_A\,C_F}\over 3} & 0 \cr
(10,\,3) & 0 & 0 & {{4\,C_A\,\left( 3 - 14\,C_F \right) }\over 3} & 0 \cr
(11,\,1) & {{4\,C_F}\over 3} & {{20\,C_F}\over 9} & {{-176\,C_F}\over {27}}
+ {{8\,C_A\,C_F\,\bar{n}_f}\over 3} & -16\,C_F \cr
(11,\,3) & {{4\,C_F}\over 9} & {{20\,C_F}\over 9} & {{944\,C_F}\over {27}} &
-24\,C_F\cr
(12,\,1) & {{4\,C_F\,\left( 2\,\bar{n}_f - 3\,n_f \right) }\over 9} &
{{16\,C_F\,\left( 2\,\bar{n}_f - 3\,n_f \right) }\over {27}} &
{{40\,C_A\,C_F}\over 3} + {{8\,C_F\,\left( 95\,\bar{n}_f - 102\,n_f \right)
}\over {81}} & {{4\,C_F\,\left( 2\,\bar{n}_f - 3\,n_f \right) }\over 3} \cr
(12,\,3) & 0 & {{4\,C_F\,\bar{n}_f}\over 3} & {{32\,C_A\,C_F}\over 3} +
{{88\,C_F\,\bar{n}_f}\over 9} & 0 \cr
}
\end{equation}
\begin{equation}
\bordermatrix{
         & 5 & 6\cr
(9,\,1) & -8\,C_F & {{C_A}\over 2} + {{10\,C_F}\over 9} \cr 
(9,\,3) & -8\,C_F & {{C_A}\over 2} + {{10\,C_F}\over 9} \cr
(10,\,1) & 0 & 1 \cr
(10,\,3) & 0 & 1 \cr
(11,\,1) & -4\,C_F & C_A + {{20\,C_F}\over 9} \cr
(11,\,3) & {{-32}\over {3\,C_A}} + {{32\,C_A}\over 3} + {{28\,C_F}\over 3}
& C_A + {{20\,C_F}\over 9} + 2\left(n_f -\frac{2}{3}\bar{n}_f\right) \cr
(12,\,1) & 4\,C_A\,C_F & 2 + {{\left(16\,C_F - 18\,C_A\right)\,\left(
2\,\bar{n}_f - 3\,n_f
\right) }\over {27}} \cr
(12,\,3) & {{52\,C_A\,C_F}\over 3} + {{8\,C_F\,\bar{n}_f}\over 3} & 2 +
{{\left(4\,C_F - C_A\right)\,\left(3\,n_f - 2\,\bar{n}_f\right) }\over 3}
}
\end{equation}
\begin{equation}
\bordermatrix{
         & 7 & 8 \cr
(9,\,1) & {{25\,C_A}\over
3} - {{88\,C_F}\over {27}} & {{-11\,C_A}\over 6} + 4\,C_F \cr 
(9,\,3) & {{-5\,C_A}\over 3} + {{688\,C_F}\over {27}} & {{-11\,C_A}\over 6}
+ 4\,C_F \cr
(10,\,1) & -{{10}\over 3} & 2 \cr
(10,\,3) & 10 & 2 \cr
(11,\,1) & {{50\,C_A}\over 3} - {{176\,C_F}\over {27}} & {{13\,C_A}\over 3}
- 4\,C_F \cr
(11,\,3) & {{944\,C_F + 126\,C_A}\over 27} +
{{44}\over 3}\,\left(n_f - \frac{2}{3}\bar{n}_f\right) & {{92\,C_F -
19\,C_A}\over 3} + 4\left(n_f-\frac{2}{3}\bar{n}_f\right) \cr
(12,\,1) & -{{20}\over 3} + {{\left(56\,C_F - 252\,C_A\right)\,\left(
2\,\bar{n}_f - 3\,n_f \right) }\over {81}} & -8 + {{5\,C_A\,\left(
2\,\bar{n}_f - 3\,n_f \right)
}\over 9} \cr
(12,\,3) & 16 + {{22\,\left(4\,C_F - C_A\right)\,\left( 3\,n_f -
2\,\bar{n}_f\right) }\over 9} & {{2\,\left(4\,C_F -
C_A\right)\,\left(3\,n_f -2\,\bar{n}_f \right) }\over 3}
}\ .
\end{equation}
\subsection{On-shell results}
\label{subsec:on-shell}
By applying the motion equations, it is now possible to give the results 
in the following on-shell basis which closely corresponds to the
symmetrization of the one used by Ciuchini et
al.~\cite{ciuchini:plb316,ciuchini:plb301},  
\begin{eqnarray}
Q^+_1 &=& \bar{Q}_1^+ - E_1^+ = \bar{O}^+_{(9,\,1)} +
\bar{O}^+_{(9,\,3)} - E_1^+\nonumber\\
Q^+_2 &=& \bar{Q}_2^+ - E_2^+ = \bar{O}^+_{(10,\,1)} +
\bar{O}^+_{(10,\,3)} - E_2^+\nonumber\\
Q^+_{3/5} &=& \bar{Q}_{3/5}^+ - E_{3/5}^+ = \bar{O}^+_{(11,\,1)} \pm
\bar{O}^+_{(11,\,3)} - E_{3/5}^+\nonumber\\
Q^+_{4/6} &=& \bar{Q}_{4/6}^+ - E_{4/6}^+ = \bar{O}^+_{(12,\,1)} \pm
\bar{O}^+_{(12,\,3)} - E_{4/6}^+\nonumber\\
Q^+_7 &=& \bar{Q}_7^+ = \bar{O}^+_5\nonumber\\
Q^+_8 &=& \bar{Q}_8^+ = \bar{O}^+_8\nonumber\\
Q^+_9 &=& \bar{Q}_9^+ - E_9^+ = \bar{O}^+_{(9,\,1)} -
\bar{O}^+_{(9,\,3)} - E_9^+\nonumber\\
Q^+_{10} &=& \bar{Q}_{10}^+ - E_{10}^+ = \bar{O}^+_{(10,\,1)} -
\bar{O}^+_{(10,\,3)} - E_{10}^+\ .
\label{eq:on-shellBasis}
\end{eqnarray}
The last two operators have been introduced to have an invertible
relation between the bases $\bar{O}^+_i$ and $Q^+_i$; as expected they
decouple from the others in the RG evolution.
The presence of the evanescent operators takes into account the
difference in the $d$-dimensional extensions, as explained in
section~\ref{subsec:thescheme}. In the basis $\bar Q_i^+$ we give the
final results for the AD matrix in the following block form: 
\begin{equation}
\hat{\gamma} = \frac{\alpha_s}{4\pi}\:\left(\begin{array}{cc}
\hat{\gamma}_{ff} & \hat{\gamma}_{fm} \cr
\hat{0} & \hat{\gamma}_{mm}
\end{array}
\right)\ .
\end{equation}
The matrices $\hat{\gamma}_{ff},\,\hat{\gamma}_{mm}$ result from the one 
loop computation and are scheme independent; they guide the RG evolution 
respectively in the four fermion and in the magnetic momentum sectors,
\begin{eqnarray}
\hat{\gamma}_{ff} &=& 
\bordermatrix{
  & 1 & 2 & 3 & 4 & 5 & 6 \cr
1 & -\frac{6}{C_A} & 6 & 0 & 0 & 0 & 0 \cr
2 &  6 & -\frac{6}{C_A} & -\frac{2}{3\,C_A} & \frac{2}{3} & -\frac{2}{3\,C_A}
& \frac{2}{3} \cr
3 & 0 & 0 & -\frac{22}{3\,C_A} & \frac{22}{3} & - \frac{4}{3\,C_A} &
\frac{4}{3} \cr
4 & 0 & 0 & \frac{2\left(9\,C_A - n_f\right)}{3\,C_A} & - \frac{2\left(9  -
n_f\,
C_A\right)}{3\,C_A} & - \frac{2\,n_f}{3\,C_A} & \frac{2\,n_f}{3} \cr
5 & 0 & 0 & 0 & 0 & \frac{6}{C_A} & - 6 \cr
6 & 0 & 0 & - \frac{2\,n_f}{3\,C_A} & \frac{2\,n_f}{3} & -
\frac{2\,n_f}{3\,C_A} &
\frac{2\left(n_f -18\,C_F\right)}{3}
}\ ,\nonumber\\
\hat{\gamma}_{mm} &=& 
\bordermatrix{
  & 7 & 8 \cr
7 & 8\,C_F & 0 \cr
8 & 8\,C_F & 16\,C_F - 4\,C_A
}\ .
\end{eqnarray}
The matrix $\gamma_{fm}$ connecting the four fermion and the magnetic
momentum sectors results from the two-loop computation and it is
scheme dependent, even at leading order, as pointed out by Ciuchini et
al.~\cite{ciuchini:plb316,ciuchini:plb301}

\begin{equation}
\hat{\gamma}_{fm} =
\bordermatrix{
  & 7 & 8 \cr
1 & 0 & 6 \cr
2 & -\frac{232\,C_F}{9} & -\frac{8\,C_A}{3} + \frac{92\,C_F}{9} \cr
3 & \frac{280\,C_F}{9} & \frac{280\,C_F}{9} + 6\,n_f \cr
4 & \frac{64\,C_A\,C_F}{3} + \frac{4\,C_F}{9}\left(27\,\bar{n}_f -
4\,n_f\right) & -4 -\frac{8\,C_A}{3}\,n_f + \frac{92\,C_F}{9}\,n_f \cr
5 & - \frac{104\,C_F}{3} &
\frac{32\,C_A}{3} - \frac{104\,C_F}{3} - 6\,n_f \cr
6 & -\frac{40\,C_A\,C_F}{3} - \frac{4\,C_F}{9}\left(4\,n_f +
27\,\bar{n}_f\right) & -8 +\frac{10\,C_A}{3} n_f - \frac{124\,C_F}{9} n_f
}\ .
\label{eq:ADMfm}
\end{equation}
\subsection{Matrix elements and scheme independence}
\label{subsec:matel}
The scheme dependence of the matrix $\hat{\gamma}_{fm}$ is compensated
in the physical amplitude by the matrix elements of the four fermion
operators.

The ``even'' contribution to the amplitude for the $b\rightarrow s\gamma$
decay has the form
\begin{equation}
<s\gamma|{\cal H}_{eff}^+|b> = \bar{C}^s_7
\left<s\gamma\left|\N{Q^+_7}\right|b\right>_0 + \sum_{n=3}^6
\bar{C}^s_{n} \left<s\gamma\left|\N{Q^+_n}\right|b\right>_1
\end{equation}
where the subscripts $0,\,1$ address the order in $\hbar$ of the matrix
element, and an analogous formula holds for the $b\rightarrow s g$
amplitude
\footnote{The bar over the coefficients reminds us that they are
derived from Eqs.~(\ref{eq:offShellCoefficients},~\ref{eq:csFromC}),
after applying motion equations: for instance $\bar{C}^s_7 = C^s_2 +
C^s_5,\,\bar{C}^s_8 = C^s_6 + C^s_8$.}.

The one loop matrix elements of the operators $Q^+_{3,\,4,\,5,\,6}$ result
from Feynman diagrams analogous to the ones listed  in
Fig.~(\ref{fig:redev})  and with a mass insertion. 
If we define our scheme preserving the naive four dimensional Fierz
simmetry they give a zero result. In fact in this case the ``open''
diagram in Fig.~(\ref{fig:redev}) with a mass insertion is
proportional to the ``closed'' one, which is set to zero by the trace.
The symmetry can be preserved by using the $HV$ scheme for $\gamma_5$ and
maintaining the four dimensional definition for the current-current
products~\cite{ciuchini:plb301}.

In other regularization schemes this is no longer true and in general
one obtains a local, finite contribution.  
This difference can be reinterpreted as a finite
correction\footnote{see, f. i. in \cite{buras:TUM-T31-50/93}, the so
called ``effective coefficients'' formalism} to the
coefficient $C^s_7$ or alternatively, according to the discussion 
at the end of Sec.~(\ref{subsec:thescheme}), to the operators. 
In our scheme the following relations hold
\begin{eqnarray}
\lefteqn{\left<s \gamma\left| \left(\begin{array}{c} \bar{O}^+_{11,n}
\cr \bar{O}^+_{12,n}
\cr \end{array}\right)
\right|b \right>_1 =} \qquad \qquad & \nonumber\\
& \left({{\left( -36 + 64\,n - 19\,{n^2} \right) }\over
{36}} \frac{c_n(0)}{n!} + {{\left( 1 - n \right) \,\left( -3 + n
\right) }\over  6}\, \frac{c_n(1)}{n!} \right) 
\left(\begin{array}{c} 1 \cr C_A \end{array}\right) \left<s
\gamma\left| \bar{O}^+_5 \right|b\right>_0 \nonumber \\ 
\lefteqn{\left<s g\left| \left(\begin{array}{c} \bar{O}^+_{11,n} \cr
\bar{O}^+_{12,n} \cr 
\end{array}\right) \right|b \right>_1 =} \qquad \qquad &\\
& \left({{\left( -36 + 64\,n - 19\,{n^2} \right) }\over
{36}} \frac{c_n(0)}{n!} + {{\left( 1 - n \right) \,\left( -3 + n
\right) }\over 6}\, \frac{c_n(1)}{n!} \right)\left(\begin{array}{c} 1 \cr 0
\end{array}\right) \left<s g\left| \bar{O}^+_8 
\right|b\right>_0 \ .\nonumber
\end{eqnarray}
Using the definitions in Eq.~(\ref{eq:on-shellBasis}) we can write
\begin{eqnarray}
\left<s\,\gamma\left|\bar{Q}^+_i\right|b\right>
&=& Z_{7 i} \left<s\,\gamma\left|\bar{Q}^+_7\right|b\right>
\qquad i=\{1 \dots 6 \} \nonumber\\
\left<s\,g\left|\bar{Q}^+_i\right|b\right>
&=& Z_{8 i} \left<s\,g\left|\bar{Q}^+_8\right|b\right>
\qquad i=\{1 \dots 6 \}
\label{eq:defZeta}
\end{eqnarray}
where following~\cite{ciuchini:plb316} we define the vectors
\begin{eqnarray}
\vec{Z}_7 &=& \left(0,0,\frac{8}{3}, \frac{8\,C_A}{3}, -\frac{2}{3},
-\frac{2\,C_A}{3}, 0, 0\right)\nonumber\\
\vec{Z}_8 &=& \left(0,0,\frac{8}{3}, 0, -\frac{2}{3}\ , 0, 0, 0\right)\ .
\label{eq:z7z8}
\end{eqnarray}
Following the steps outlined in Sec.~\ref{sec:strategy} we perform the
finite renormalizations needed to compare the results with the ones
obtained in the HV scheme.
We match the renormalization schemes by defining a non-minimal subtraction
\begin{eqnarray}
\Np{\bar{Q}^+_i} &=& \left(\hat{1} + \delta\hat{F}\right)_{i\,j}
\N{\bar{Q}^+_j}\nonumber\\
\delta\hat{F}_{i\,j} &=& \left\{\begin{array}{lr} - Z_{j\,i} &
j = \left\{7,\,8\right\} \cr
0 & j = \left\{1,\dots\,6\right\}
\end{array} \right.
\label{eq:finitematching}
\end{eqnarray}
which sets to zero the matrix elements in Eq.~(\ref{eq:defZeta}), as they 
are in the HV scheme.

According to the formula in Eq.~(\ref{eq:RGfinite}) the AD matrix is
modified as follows 
\begin{equation}
\hat{\gamma}^\prime = \left(\hat{1} + \delta\hat{F}\right) \hat{\gamma}
\left(\hat{1} + \delta\hat{F}\right)^{-1}\ .
\end{equation}
Using the results in Eq.~(\ref{eq:z7z8}), one easily obtains the AD
matrix in the HV scheme
\begin{eqnarray}
\hat{\gamma}^\prime_{mf} =
\bordermatrix{
  & 7 & 8 \cr
1 & 0 & 6 \cr
2 & -\frac{208\,C_F}{9} & \frac{116\,C_F}{9} - 4\,C_A \cr
3 & \frac{232\,C_F}{9} & \frac{232\,C_F}{9} - 8\,C_A + 6 n_f \cr
4 & \frac{8\,C_F}{9}\,n_f + 12\,C_F\,\bar{n}_f & 12 +
\left(\frac{116\,C_F}{9} - 4\,C_A\right) n_f \cr
5 & -16\,C_F & 4\,C_A - 16\,C_F - 6\,n_f\cr
6 & \frac{8\,C_F}{9}\,n_f - 12\,C_F\,\bar{n}_f & -8 + \left(2\,C_A -
\frac{100\,C_F}{9}\right) n_f
}
\label{eq:ADMHV}
\end{eqnarray}
which coincides with the result in~\cite{ciuchini:plb316}.

Recall that in Eq.~(\ref{eq:finitematching}) we have imposed that the
renormalized operators in the symmetrized scheme have the same matrix
elements as in the HV scheme: moreover, at LO the coefficients at
$\mu\simeq M_W$, relevant for the $b\rightarrow s\gamma$, are unaffected
by the scheme redefinition, hence we conclude that the two schemes are
completely equivalent. In other words the physical amplitude is now
determined by the evolved coefficients (in the ``symmetrized'' scheme
supplemented with the finite renormalization) and by the operators
renormalized in the HV scheme.

At NLO it will be necessary a two loop matching of the schemes, but the
three loop computations will be performed only in the symmetrized scheme.
%
%
\section{Checks of the two loop computation}
\label{sec:checks}
In this section we want to discuss the redundancy of the computation, 
which has been used in this work as a check of the computer codes and of 
the method itself; this can be of great help to reduce the complexity
of a future NLO computation.

\subsubsection{Fierz Identities}

The generalized Fierz Identity~\cite{avdeev,kennedy} listed in 
App.~\ref{app:dirac}, Eq.~\ref{fierz:def}, allow to relate ``bare'' 
graphs with a closed loop, listed in
Figs.~(\ref{fig:mix21},\ref{fig:mix21a}), with the  ones 
with an open loop, listed in Figs.~(\ref{fig:mix22},\ref{fig:mix22a}).

In other words, let $Closed\left(n\right)$ be the result of a graph with a
closed loop, barring the sign coming from the fermionic loop and any color
or flavor structure, and let $Open\left(n\right)$ be the corresponding graph
with an open loop, having exactly the same ``topology'': corresponding
graphs in Figs.~(\ref{fig:mix21},\ref{fig:mix21a})
and~(\ref{fig:mix22},\ref{fig:mix22a}) are examples of such pairs and
the index $n$ addresses the order of the $\gbb{n}$ operator inserted. 

Then the following relation holds:
\begin{equation}
Open\left(n\right) = \frac{1}{\mathrm{Tr}\left(1\right)}*\sum_{j=1}^\infty
\frac{1}{j!}\left(-1\right)^{\frac{j\,(j-1)}{2}} f_{n\,j}
Closed\left(j\right).
\label{eq:checkFierz}
\end{equation}
By expanding both members in powers of $\varepsilon$~\footnote{It is
 crucial to expand also the $f_{n\,j}$ function, which depends on
$d$.} and noting that on the right hand side only a finite number of
terms can contribute, at a given perturbative order, one gets
relations among the double and the single poles. Therefore the
computation of  one of the two sets of massless and massive graphs
would be enough (as far as  the ``bare'' part is concerned),
 provided one maintains $n$
arbitrary. Note that not only odd values of $n$ are needed
 and that  is the reason why we have kept all values of $n$,
despite the fact that only odd values of $n$ appear in the
anomalous dimension matrices. This trick is not straightforwardly
applicable when dealing with a counterterm graph if  the
four fermion vertex belongs to one of the divergent subgraphs
considered. In fact in this case the subtraction recipe imposes the
substitution of the subgraphs with the respective pole parts;
this
operation spoils the Fierz identity, since it is a ``projection'' on the
physical $4$-dimensional space that does not commute with the
transformation in Eq.~(\ref{eq:checkFierz}).

\subsubsection{Charge Conjugation}

Let us recall the properties of the charge conjugation operator in
euclidean notation

\begin{eqnarray}
C^{-1} \gamma_\mu^T C &=& - \gamma_\mu\,\nonumber\\
C^T =  C^\dagger &=& - C = C^{-1}\,\nonumber\\
C C^\dagger = C^\dagger C &=& - C^2 = 1\,\\
C^{-1} \gb{n} C &=& \left(-1\right)^n \gb{n}\,\nonumber\\
C^{-1} \left(\gb{n}\right)^T C &=&
\left(-1\right)^{\frac{n\,\left(n+1\right)}{2}} \gb{n}\ .\nonumber
\end{eqnarray}

The last pair of relations is trivially derived by inserting $C C^{-1}$
factors in between $\gamma$ matrices and using
\begin{equation}
\left(\gb{n}\right)^T = 
\left(-1\right)^{\frac{n\,\left(n-1\right)}{2}}\gb{n}\ .
\end{equation}

\begin{figure}
\begin{center}
\leavevmode
\epsfbox{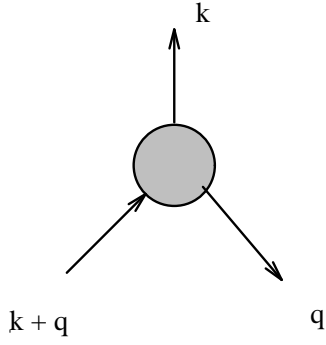}
\end{center}
\caption{Naming of external momenta in the two loop diagrams}
\protect\label{fig:momFlow}
\end{figure}

We will take as reference Fig.~(\ref{fig:momFlow}) for the flow of momenta and
Figs.~(\ref{fig:mix21},~\ref{fig:mix21a},~\ref{fig:mix22},~\ref{fig:mix22a})
for the naming of graphs.

\paragraph{Open loops}

Sandwiching the graphs in Fig.~(\ref{fig:mix21}) between $C^{-1}$ and
$C$, it is easy to show that the following relations hold
\begin{eqnarray}
C^{-1} A^T\left(k,\,q\right) C &=& - A\left(k,\,-k-q\right)\nonumber\\
C^{-1} B^T\left(k,\,q\right) C &=& - D\left(k,\,-k-q\right)\nonumber\\
C^{-1} C^T\left(k,\,q\right) C &=& - E\left(k,\,-k-q\right)\nonumber\\
C^{-1} F^T\left(k,\,q\right) C &=& - F\left(k,\,-k-q\right)\nonumber\\
C^{-1} G^T\left(k,\,q\right) C &=& - H\left(k,\,-k-q\right)\nonumber\\
C^{-1} I^T\left(k,\,q\right) C &=& - J\left(k,\,-k-q\right)\nonumber\\
C^{-1} K^T\left(k,\,q\right) C &=& - L\left(k,\,-k-q\right)\nonumber\\
C^{-1} M^T\left(k,\,q\right) C &=& - M\left(k,\,-k-q\right)
\end{eqnarray}
Therefore most graphs are paired; for $A,\,F,\,M$ the relation is
trivially satisfied. It is also worth remembering that the relations among
graphs with the trilinear gluon vertex take into account its antisymmetry
properties, once the same color factor is factorized for the three
$K,\,L,\,M$ graphs.

\paragraph{Closed loops}

In this case, see Figs.~(\ref{fig:mix22},~\ref{fig:mix22a}), we have two
fermionic lines, one traced and the other connected to the ``in'' and
``out'' states. Hence two sets of relations are possible.

By exploiting $C$ invariance on the external line, one gets the relations
\begin{eqnarray}
C^{-1} B^T\left(k,\,q\right) C &=&
\left(-1\right)^{\frac{n\,\left(n-1\right)}{2}}
E\left(k,\,-k-q\right)\nonumber\\ 
C^{-1} D^T\left(k,\,q\right) C &=&
\left(-1\right)^{\frac{n\,\left(n-1\right)}{2}}
C\left(k,\,-k-q\right)\nonumber\\ 
C^{-1} F^T\left(k,\,q\right) C &=&
\left(-1\right)^{\frac{n\,\left(n+1\right)}{2}}
F\left(k,\,-k-q\right)\nonumber\\
C^{-1} I^T\left(k,\,q\right) C &=&
\left(-1\right)^{\frac{n\,\left(n+1\right)}{2}}
J\left(k,\,-k-q\right)\nonumber\\
C^{-1} K^T\left(k,\,q\right) C &=&
- \left(-1\right)^{\frac{n\,\left(n-1\right)}{2}}
L\left(k,\,-k-q\right)
\end{eqnarray}
while inserting a $C C^{-1}$ factor in the fermion loop, one is able to
turn a clockwise loop in an anti-clockwise one, thus obtaining the
relations\footnote{In the last relation a definite convention for the
color factors is assumed.}
\begin{eqnarray}
B\left(k,\,q\right) &=& \left(-1\right)^{\frac{n\,\left(n+1\right)}{2}}
C\left(k,\,q\right)\nonumber\\
D\left(k,\,q\right) &=& \left(-1\right)^{\frac{n\,\left(n+1\right)}{2}}
E\left(k,\,q\right)\nonumber\\
G\left(k,\,q\right) &=& \left(-1\right)^{\frac{n\,\left(n-1\right)}{2}}
H\left(k,\,q\right)\nonumber\\
M\left(k,\,q\right) &=& - \left(-1\right)^{\frac{n\,\left(n+1\right)}{2}}
M\left(k,\,q\right)
\end{eqnarray}

By using the Fierz Identities and the relations coming from the charge 
conjugation, one reduces the computation of $26$ graphs to that of at 
most $10$. 

\subsubsection{Non local poles cancellation.}

The total counterterm part of a two loop graphs exhibits a
double pole with a coefficient twice the double pole in the ``bare''
graphs, thus ensuring the cancellation of non-local poles. This has been 
verified, together with the explicit cancellation of the logarithms,
which is afurther test of the computer codes.

\subsubsection{Gauge Invariance}

We have verified that the gauge invariant basis off-shell,
Eq.~(\ref{eq:off-shellBasis}), is a complete basis;
 therefore all the results can be expressed in an explicit gauge
invariant way, even off-shell, thanks to the background gauge-fixing.

The gauge invariance reflects itself in the fact that the Ward Identities
for the vertex
\begin{equation}
k_\mu V^{b\rightarrow s\gamma}_\mu\left(k+q,\,q\right) \propto
S_s\left(q\right) - S_b\left(k+q\right)
\label{eq:ward}
\end{equation}
apply separately for different classes of graphs; therefore the groups
\begin{itemize}
\item \{A, G, H\}
\item \{B, C, J\}
\item \{D, E, I\}
\item \{F\}
\item \{K, L, M\}
\end{itemize}
can be individually checked.

\subsubsection{Gauge Independence}

All the graphs have been evaluated with an arbitrary background gauge
fixing, depending on a parameter $\xi$. It is known that the resulting
anomalous dimensions, on-shell, are independent from this gauge parameter:
gauge dependence is possible only for operators proportional to the motion
equations, which are zero on-shell, as we have seen in 
Sec.~\ref{subsec:oneloop}.
The cancellation is effective among classes of graphs which are separately
gauge independent:
it is also worth noting that the cancellation holds separately for the 
two Casimir invariants, $C_A$ and $C_F$.
%
%
\section{Conclusions}
\label{sec:conclusions}
In this work we have detailed the computation of the leading logarithmic 
corrections to the $b\rightarrow s\gamma$ decay, presented 
in~\cite{CCRV:letter}, which confirms the results of Ciuchini et 
al.~\cite{ciuchini:plb316,ciuchini:rome93/973}.

The application to this problem of the method 
introduced in~\cite{curci:ricc:prd47} appears successful and the 
practical implementation provides relevant simplifications over the
only other method to our knowledge free of ambiguities, the HV scheme.

The relevance of the next to leading computation has been stressed by 
several authors and we refer to their works for extended discussions, 
concerning the possibility to test more stringently the Standard 
Model itself~\cite{buras:TUM-T31-50/93} or to restrict the parameter 
space for extended models, such as the Minimal Supersymmetric Standard
Model~(see f.i.~\cite{borzumati:desy93-090,bertolini:vissani:94}). 

We do not think that the NDR and DRED schemes, although 
shown to be consistent up to the two loop 
level~\cite{buras:mpi-pae56-91,ciuchini:plb301}, are suited for the 
three loop computation of the next to leading corrections. Moreover, our 
opinion is that even if some tricks were devised to avoid the ambiguities 
known to be present in the NDR and DRED 
schemes~\cite{avdeev:chochia:vladimirov}, the approach we propose would 
be neverteless simpler thanks to the unified treatment of the evanescent 
and relevant operators.
\appendix
%
%
\section{Renormalization Group formulas}
\label{app:rg}
In this appendix we give some reference formulas for dimensional 
regularization and the renormalization group: they express well known 
results, but we think useful to give them in a somewhat condensed form.

We consider a field theory describing a certain set of fields $\{\phi\}$, 
regularized by continuation in the number of dimensions
($d = 4 - 2\varepsilon$), and defined by the action
\begin{equation}
S = \int \d^d z \left(\mu^2\right)^{-\varepsilon}\sum_i\left(g_i + 
P_i\right) {\cal L}_i\left(z\right)
\label{eq:actionDR}\ ;
\end{equation}
the counterterms ${P}$ renormalize the different couplings $g_i$, 
defining the $\varepsilon \rightarrow 0$ limit.
Given a set of local operators $\left\{G\right\}$, which is a complete
basis for the  renormalization, they are defined at quantum level by
the normal product
\begin{equation}
\N{G_i} = \N{k_i O_i} = \sum_j k_i M_{i\,j} O_j\ ,
\label{eq:defN}
\end{equation}
where the $k_i$ are the explicit couplings (for instance, $g_s$ or $m_b$ in 
our case) and the $M_{i,\,j} = \delta_{i,\,j} + O\left(\hbar\right)$ are
the counterterms.

The following RG equation in $d$ dimensions holds,
\begin{equation}
\left\{\left[\mu\frac{\partial}{\partial\,\mu} + 
\sum_l\left(\beta_l +  2\varepsilon g_l\right) 
\frac{\partial}{\partial g_l} - \gamma_\phi \frac{n}{2}\right] 
\delta_{i,\,j} + \gamma_{i,\,j}\right\}
\left<0\left|\N{G_j}\phi_1\dots\phi_n\right|0\right>_{\mathrm{1PI}} = 0
\label{eq:RG}
\end{equation}
where
\begin{equation}
	\gamma_\phi = -\sum_L 2 L P_\phi^{L,1}
\end{equation}
is the field anomalous dimension, that is the $\beta$ function associated 
with the kinetic term in the zero mass limit, while 
\begin{equation}
	\beta_l = \sum_L 2 L P_l^{L,1} - \gamma_\phi\frac{\rho^l}{2}g_l\ ,
\end{equation}
is the $\beta$ function associated with the coupling $g_l$ in the 
interaction part of the action and
\begin{equation}
\gamma_{i,\,j} = \sum_L 2 L M_{i,\,j}^{L,\,1}\frac{k_i}{k_j} + 
\left(\frac{\gamma_\phi}{2} \nu^i - \sum_l 
\beta_l\frac{\partial\log{k_i}}{\partial g_l}\right) \delta_{i,\,j}\ ,
\label{eq:RGfun}
\end{equation}
is the anomalous dimension matrix of the operators $\N{G_i}$.
The sum is over $L$, the number of loops, and the symbol $\rho_l$ stands 
for the number of elementary fields entering the interaction term ${\cal 
L}_l$, while $\nu^i$ is the analogous quantity for the operator 
$\N{G_i}$.
The existence of the $d\rightarrow 4$ limit of the RG equation imposes a 
set of recursive relations which expresses the well known fact that at 
a certain order in the loop expansion all the pole parts except the 
simple pole are predicted by the results of the computation at the 
preceding orders,
\begin{eqnarray}
\sum_L 2 L M_{i,\,k}^{L,p+1} & = & \sum_{L,L^\prime} 2 
L M_{i,\,j}^{L,\,1} M_{j,\,k}^{L^\prime,\,p} \nonumber\\
 &\phantom{=}& + \sum_{L,\,l}\beta_l\frac{\partial M_{i,\,k}^{L,\,p}}{\partial 
 g_l} + \frac{\gamma_\phi}{2} \sum_L \left(\nu^i - \nu^j\right) 
 M_{i,\,k}^{L,\,p}\nonumber\\
 \sum_L 2 L P_k^{L,\,p+1} &=& \sum_{L,\,l} \left(\beta_l 
 \frac{\partial}{\partial g_l} - \frac{\gamma_\phi}{2} \rho^k\right) 
 P_k^{L,\,p}\ .
\label{eq:recursivePoles}
\end{eqnarray}
Altough the operator basis is in general infinite, in $d$ dimensions, at 
a fixed order in perturbation theory only a limited set of operators can 
contribute to the physics. Moreover, the evanescent operators can be 
``eliminated'', in the sense that the $d\rightarrow 4$ limit of the RG 
equation can be defined, which takes into account the effect of 
evanescent operators, through a redefinition of the anomalous dimension 
matrix.
\subsection{Reduction of evanescents}
\label{subsec:redev}
Let us separate the basis $\{G\}$ in relevant $\{R\}$ and evanescent 
$\{E\}$ operators: the relevant ones have a non-zero classical 
limit, while the evanescent operators contribute to matrix elements only 
at the quantum level. A reduction formula
\begin{equation}
\N{\vec{E}} = \hat{r}_{E,\,R} \N{\vec{R}}
\label{eq:reductionFormula}
\end{equation}
relates perturbatively evanescent and relevant operators and the 
$\varepsilon\rightarrow 0$ limit is modified: the coupled equations in
Eq.~(\ref{eq:RG}) can be written in a schematic form as
\begin{eqnarray}
{\cal D} \N{\vec{R}} + \hat{\gamma}_{R,\,R^\prime} \N{\vec{R}^\prime} +  
\hat{\gamma}_{R,\,E} \N{\vec{E}} & = & 0\ , \nonumber\\
{\cal D} \N{\vec{E}} + 
\hat{\gamma}_{E,\,E^\prime} \N{\vec{E}^\prime} & = & 0\ .
\label{eq:RGwithEvanescent}
\end{eqnarray}
By using Eq.~(\ref{eq:reductionFormula}) the evanescent operators can be 
eliminated and  the evolution of the relevant operators written only in terms 
of themselves
\begin{eqnarray}
\left[{\cal D} \hat{\delta} + \hat{\gamma}^\prime\right] \N{\vec{R}} &
= & 0\nonumber\\
\hat{\gamma}^\prime_{R\,R^\prime} &=& \hat{\gamma}_{R,\,R^\prime} +
\hat{\gamma}_{R,\,E} \hat{r}_{E,\,R^\prime}\ ,
\label{eq:RGreduced}
\end{eqnarray}
while the second relation in Eq.~(\ref{eq:RGwithEvanescent}) gives a
consistency condition
\begin{equation}
\left[\left({\cal D} \hat{r}_{E,\,R}\right) -
\hat{r}_{E,\,R^\prime}\left(\hat{\gamma}_{R^\prime,\,R} +
\hat{\gamma}_{R^\prime,\,E^\prime} \hat{r}_{E^\prime,\,R}\right) +
\hat{\gamma}_{E,\,E^\prime} \hat{r}_{E^\prime,\,R}\right] \N{\vec{R}}
= 0\ .
\label{eq:RGconsistency}
\end{equation}
In this way we have directly projected the RG equation in $4$ dimensions.
\subsection{Non minimal prescription}
\label{subsec:nonminimal}
Let us compare with a different approach
by using finite renormalizations in
order to impose that the evanescent operators decouple from the evolution
of relevant ones. This means choosing a non-minimal subtraction scheme
\begin{equation}
\Np{\vec{E}} = \N{\vec{E}} - \hat{r}_{E,\,R} \N{\vec{R}}\ ,
\label{eq:RenormalizedEvanescent}
\end{equation}
thus setting to zero the matrix elements of evanescent operators, in the 
$d\rightarrow 4$ limit.

It is trivial to demonstrate that, given
\begin{equation}
\Np{G_i} = F_{i\,j} \N{G_j}
\label{eq:finiteRen}
\end{equation}
the anomalous dimension matrix is modified as follows:
\begin{equation}
\hat{\gamma}^\prime = \hat{F}\hat{\gamma}\hat{F}^{-1} - \left(\beta_l
\frac{\partial \hat{F}}{\partial g_l} - 2\varepsilon \hat{L}
\hat{F}\right) \hat{F}^{-1}\ ,
\label{eq:RGfinite}
\end{equation}
where $\hat{\gamma}$ is the anomalous dimension matrix in a minimal scheme
and $\hat{L}$ is the loop-counting operator:
writing
\[
\hat{F} = \hat{1} - \hat{r}
\]
one directly finds
\begin{equation}
\hat{\gamma}^\prime_{R,\,R^\prime} = \hat{\gamma}_{R,\,R^\prime} + 
\hat{\gamma}_{R,\,E} \hat{r}_{E,\,R^\prime},
\end{equation}
that is, the relevant anomalous dimension matrix is modified exactly as in
Eq.~(\ref{eq:RGreduced}).

The other submatrices
\begin{eqnarray}
\hat{\gamma}^\prime_{R,\,E} &=& \hat{\gamma}_{R,\,E}\nonumber\\
\hat{\gamma}^\prime_{E,\,E^\prime} &=& \hat{\gamma}_{E,\,E^\prime} -
\hat{r}_{E,\,R}\hat{\gamma}_{R,\,E^\prime}
\end{eqnarray}
are  irrelevant, since the evanescent operators are now
subtracted to give zero contribution to matrix elements. Note that in 
the non minimal prescription the submatrix $\hat{\gamma}_{E,\,R}$ receives
a contribution
\begin{equation}
\hat{\gamma}^\prime_{E,\,R} =  -
\hat{r}_{E,\,R^\prime}\left(\hat{\gamma}_{R^\prime,\,R} +
\hat{\gamma}_{R^\prime,\,E^\prime} \hat{r}_{E^\prime,\,R}\right) +
\hat{\gamma}_{E,\,E^\prime} \hat{r}_{E^\prime,\,R}\ ,
\end{equation}
which is set to zero by the consistency condition in 
Eq.~(\ref{eq:RGconsistency}).

\subsection{Reference for QCD}
\label{subsec:refQCD}

The relevant RG equation for QCD at leading logarithmic order is given, 
in simplified notation, by
\begin{equation}
\left({\cal D} \delta_{i,\,j} + \gamma_{i,\,j}\right) \N{G_i} = 0
\end{equation}
\begin{equation}
{\cal D} = \mu\frac{\partial}{\partial \mu} + \beta_{QCD}
\frac{\partial}{\partial g_S} + \gamma_{m_b}\, m_b \frac{\partial}{\partial
m_b} + \gamma_\xi \frac{\partial}{\partial \xi}\ ,
\end{equation}
and the $\beta$ and $\gamma$ functions needed are given in the background
field gauge
\begin{eqnarray}
\gamma_f &=& \frac{\alpha_s}{2\pi}\xi\,C_F\ ,\\
\gamma_g &=& - \frac{\alpha_s}{2\pi}\,\left(\frac{11}{3}\,C_A - \frac{2}{3}
n_f\right)\ , \\
\gamma_{m_b} &=& - \frac{\alpha_s}{2\,\pi}\,3\,C_F\ ,\\
\beta_{QCD} &=& - g_s \frac{\alpha_s}{4\pi} \,\left(\frac{11}{3} C_A -
\frac{2}{3} n_f\right)\ .
\end{eqnarray}
The anomalous dimension  of the gauge fixing term $\gamma_\xi$ does
not contribute at this order in perturbation theory.
%
%
\section{Dirac Algebra}
\label{app:dirac}

We collect here some useful Dirac algebra definitions and identities,
mostly taken from~\cite{tHooftVeltman,avdeev,kennedy}.
\begin{eqnarray}
\left\{\gamma_\mu,\;\gamma_\nu\right\} &=& 2
\delta_{\mu\nu}\qquad\qquad\gamma_5 = - \gamma_1\gamma_2\gamma_3\gamma_4 =
- \frac{1}{4!}
\varepsilon_{\mu\nu\rho\sigma}\gamma_\mu\gamma_\nu\gamma_\rho\gamma_\sigma
\nonumber\\
\varepsilon_{1234} &=& + 1\nonumber\\
\gamma_{\mu\nu\rho} &\equiv& \frac{1}{2}\left(\gamma_\mu\gamma_\nu\gamma_\rho
- \gamma_\rho\gamma_\nu\gamma_\mu\right) = -
\varepsilon_{\mu\nu\rho\sigma}\gamma_5\gamma_\sigma\ . 
\end{eqnarray}
Note also that we define
\begin{equation}
P_{L/R} = \frac{1}{2}\left(1 \mp \gamma_5\right)\qquad\qquad
\gamma_\mu^{L/R} = \gamma_\mu P_{L/R}\ .
\end{equation}
We have used the completely antisymmetric products of
$\gamma$ matrices as a complete basis of the Dirac algebra in $d$ dimensions
\begin{equation}
\gamma_{\mu_1\mu_2\dots\mu_n} =
\frac{1}{n!}\sum_p \left(-1\right)^p
\gamma_{\mu_1}\gamma_{\mu_2}\dots\gamma_{\mu_n}\ ,
\end{equation}
and it is useful to introduce, following Avdeev~\cite{avdeev} the
antisymmetrization operator
\begin{equation}
g_{\mu\nu}^{m} = \frac{1}{m!}\sum_{\pi\in s_m}
\left(-1\right)^{p\left(\pi\right)} \delta_{\mu_1\nu_{\pi(1)}}\dots
\delta_{\mu_m\nu_{\pi(m)}}
\end{equation}
which is a projection
\begin{equation}
g_{\mu\nu}^{m} g_{\nu\rho}^m = g_{\mu\rho}^m
\end{equation}
and its trace is given by
\begin{equation}
g_{\mu\mu}^m = \left(\begin{array}{c} d \\
m\end{array}\right)\equiv\frac{1}{m!}\prod_{i=1}^m\left(d - m + i\right)\ .
\end{equation}
We give some  identities valid in the limit of
$d\rightarrow 4$:
\begin{eqnarray}
\gamma_{\mu_1\mu_2\dots\mu_n} &=& 0\qquad n > 4\nonumber\\
\gamma_{\mu_1\mu_2\mu_3\mu_4} &=& - \gamma_5\nonumber\\
\gamma_{\mu_1\mu_2\mu_3} &=& - \varepsilon_{\mu_1\mu_2\mu_3\rho}
\gamma_5 \gamma_\rho\nonumber\\
\frac{1}{3!}\gamma_{\mu\nu\rho} \otimes \gamma_{\mu\nu\rho} &=&
\gamma_\mu\gamma^5\otimes \gamma_\mu\gamma^5\ .
\end{eqnarray}
In $d$ dimensions useful relations are 
\begin{eqnarray}
\gamma_\mu\gamma_\nu\gamma_\rho\otimes\gamma_\mu\gamma_\nu\gamma_\rho &=&
\gamma_{\mu\nu\rho}\otimes\gamma_{\mu\nu\rho} + \left(3
d-2\right)\gamma_\mu\otimes\gamma_\mu\ ,\nonumber\\
\gamma_\mu\gamma_\nu\gamma_\rho\otimes\gamma_\rho\gamma_\nu\gamma_\mu &=&
- \gamma_{\mu\nu\rho}\otimes\gamma_{\mu\nu\rho} + \left(3
d-2\right)\gamma_\mu\otimes\gamma_\mu\ .
\end{eqnarray}
The projection on the canonical basis of any product of gamma matrices can
be obtained by using recursive relations
\begin{eqnarray}
\gamma_\nu \gamma_{\mu_1,\mu_2\dots,\mu_n} &=& \gamma_{\nu,\mu_1\dots,\mu_n}
- \sum_i \left(-1\right)^i \delta_{\nu\mu_i}
\gamma_{\mu_1\dots\mu_{i-1},\mu_{i+1}\dots,\mu_n}\ ,\nonumber\\
\gamma_{\mu_1,\mu_2\dots,\mu_n}\gamma_\nu &=& \gamma_{\mu_1\dots,\mu_n,\nu} +
\sum_i \left(-1\right)^{n-i} \delta_{\nu\mu_i}
\gamma_{\mu_1\dots\mu_{i-1},\mu_{i+1}\dots,\mu_n}\ .
\end{eqnarray}
The following trace identity holds:
\begin{equation}
\mathrm{Tr}\left(\gb{m}\gb{n}\right) =
n! \delta_{m\,n} s_m g_{\mu\nu}^m \mathrm{Tr}\left(1\right)
\end{equation}
where for convenience we have defined
\begin{equation}
s_n = \left(-1\right)^{\frac{n (n-1)}{2}}\ .
\end{equation}
Some contraction identities on Dirac matrices that belong to the canonical
basis are now listed
\begin{eqnarray}
\gamma_{\mu_1\mu_2\dots\mu_n}\gamma_{\mu_1} &=&
\left(-1\right)^{n-1}\left(d-n+1\right)\gamma_{\mu_2\dots\mu_n}\nonumber\\
\gamma_{\mu_1\mu_2\dots\mu_n}\gamma_{\mu_1\mu_2\dots\mu_n} &=& c_n\nonumber\\
c_n &=& s_n \prod_{i=0}^{n-1}\left(d-i\right)\ .
\end{eqnarray}
In the text the $c_n$ have been expressed as a series in $\varepsilon$
defining
\begin{equation}
c_n = c_n(0) +\varepsilon\,c_n(1) +\varepsilon^2 c_n(2) +O(\varepsilon^3)\ .
\end{equation}
We also found useful the following commutation identities
\begin{eqnarray}
\left\{\gamma_{\mu_1,\mu_2,\dots,\mu_n},\ \gamma_\nu\right\} &=& 2
\gamma_{\mu_1,\mu_2,\dots,\mu_n,\nu}\qquad n\ \mathrm{even}\nonumber\\
\left[\gamma_{\mu_1,\mu_2,\dots,\mu_n},\ \gamma_\nu\right] &=& 2
\gamma_{\mu_1,\mu_2,\dots,\mu_n,\nu}\qquad n\ \mathrm{odd}\ .
\end{eqnarray}
Many relations depend only on the number of indexes $n$: addressing with
$\gb{n}$ a completely antisymmetric tensor in $n$ indexes,
one proves (inductively) that
\begin{equation}
\gamma_\mu \gb{n}\gamma_\mu = \left(-1\right)^n \left(d
- 2\,n\right) \gb{n}
\end{equation}
\begin{eqnarray}
\label{nmun}
\gb{n} \gamma_\mu \gb{n} &=&
s_n\left(2 n - d\right)\prod_{i=1}^{n-1}\left(i-d\right)
\gamma_\mu\qquad{n \geq 2}\nonumber\\
&=& \left(-1\right)^n \frac{d - 2 n}{d} c_n
\end{eqnarray}
\begin{eqnarray}
\gb{n} \gamma_\mu\gamma_\nu\gb{n} &=&
\frac{1}{d\left(d-1\right)}\left[\left(d^2 - \left(d-2 n\right)^2\right)
\delta_{\mu\nu} + \left(\left(d- 2 n\right)^2 - d\right)
\gamma_\mu\gamma_\nu\right] \gb{n} \gb{n}\nonumber\\
\gb{m} \gb{n} \gb{m} &=& f_{m\,n} \gb{n}
\end{eqnarray}
\begin{equation}
f_{m\,n} = m!\,s_n s_{m+n}\sum_{l=0}^{\min(m,n)}
\left(-1\right)^l\left(\begin{array}{c} n \\ l \end{array}\right)
\left(\begin{array}{c} d - n \\ m - l \end{array}\right)\ .
\label{f:def}
\end{equation}
In Eq.~(\ref{nmun}) the summation over repeated indices is understood; note
also in Eq.~(\ref{f:def}) the difference from the definition of $f$
in~\cite{avdeev}.

As for $c_n$ we write the series expansion of $f$ as
\[
f_{m\,n} = f_{m\,n}(0) + \varepsilon f_{m\,n}(1) + O(\varepsilon^2)\ .
\]
The $f_{m\,n}$ coefficients obey the consistency condition
\begin{equation}
\delta_{m,k}\left(\mathrm{Tr}\left(1\right)\right)^2 = \sum_{n=0}^\infty
\frac{s_n\,s_k}{m!\,n!}f_{m\,n} f_{n\,k}\ .
\label{f:consistency}
\end{equation}
Let us report an observation of Avdeev about the use of Fierz transformations:
first of all, note that the following relation holds
\begin{equation}
\gb{m}\otimes\gb{m} = \frac{1}{\mathrm{Tr}\left(1\right)}\sum_{n=0}^\infty
\frac{s_n}{n!} f_{m\,n}\left(\gb{n}\otimes\gb{n}\right)_F
\label{fierz:def}
\end{equation}
which, together with Eq.~(\ref{f:consistency}) and the consequences
\begin{equation}
f_{0\,n} = 1\qquad\mathrm{and}\qquad f_{n\,0} = n!\,s_n \left(\begin{array}{c}
d \\ n\end{array}\right)
\end{equation}
imposes the condition
\begin{equation}
\left(\mathrm{Tr}\left(1\right)\right)^2 = \sum_{l=0}^\infty
\left(\begin{array}{c}
d \\ l\end{array}\right) = 2^d\ .
\end{equation}
It follows that Fierz transformations~(\ref{fierz:def}) are inconsistent
with the usual choice $\mathrm{Tr}\left(1\right) = 4$.
On the other hand, as the values of $f_{m\,n}$ follow from purely
combinatorial relations in Eq.~(\ref{f:def}), the contraction identities remain
valid: furthermore, in Eq.~(\ref{eq:checkFierz}) the
$\mathrm{Tr}\left(1\right)$
cancels and the use of the Fierz identity is legitimate even with
$\mathrm{Tr}\left(1\right) = 4$.

We have used in the computation reduction identities of the following form
\begin{eqnarray}
\label{fierz:red:fund}
\gamma_\mu \gb{n} \otimes \gamma_\mu \gb{n} &=&
\gb{n+1}\otimes\gb{n+1} + n \left(d - n + 1\right)\gb{n-1}\otimes
\gb{n-1}\ ,\nonumber\\ 
\gb{n}\gamma_\mu \otimes \gb{n}\gamma_\mu &=&
\gb{n+1}\otimes\gb{n+1} + n \left(d - n + 1\right)\gb{n-1}\otimes
\gb{n-1}\ ,\\
\gamma_\mu \gb{n} \otimes \gb{n} \gamma_\mu &=&
\left(-1\right)^n\left(\gb{n+1}\otimes\gb{n+1} - n \left(d - n +
1\right)\gb{n-1}\otimes \gb{n-1}\right)\ .\nonumber
\end{eqnarray}
The relations in Eq.~(\ref{fierz:red:fund}) are fundamental and allow to
reduce recursively any product of the form
\begin{equation}
\gamma_{\mu_1}\dots\gamma_{\mu_k} \gb{n}
\gamma_{\mu_{k+1}}\dots\gamma_{\mu_m} \otimes
\gamma_{\mu_{\pi(1)}}\dots\gamma_{\mu_{\pi(h)}}\gb{n}
\gamma_{\mu_{\pi(h+1)}}\dots\gamma_{\mu_{\pi(m)}}
\label{fierz:red:gen}
\end{equation}
where $\pi\in S_m$ is a permutation of indexes.
%
%

%
\end{document}